\documentclass[twocolumn,showpacs,preprintnumbers,amsmath,amssymb]{revtex4}

\usepackage{graphicx}
\usepackage{dcolumn}
\usepackage{bm}

\begin{document}


\title{ELECTROMAGNETIC WAVE PROPAGATION IN GLHUA INVISIBLE SPHERE\\
             BY GL NO SCATERING FULL WAVE MODELING AND INVERSION
 }

\author{Jianhua Li}
 \altaffiliation[Also at ]{GL Geophysical Laboratory, USA, glhua@glgeo.com}
\author{  Feng Xie, Lee Xie, Ganquan Xie}%
\affiliation{%
GL Geophysical Laboratory, USA
}%

\hfill\break
\author{Ganquan Xie}
\affiliation{
Chinese Dayuling Supercomputational Sciences Center, China
}%

\date{December 26,2016}

\begin{abstract}
Using GL no scattering full wave modeling and inversion, we create a GLHUA pre cloak electromagnetic (EM) material in the virtual sphere that makes the sphere is invisible. The invisible sphere is called GLHUA sphere. In GLHUA sphere, the Pre cloak relative parameter is not less than 1; the parameters and their derivative are continuous across the boundary r=R2 and the parameters are going to infinity at origin r=0. The phase velocity of EM wave in the sphere is less than light speed and going to zero at origin. The EM wave field excited in the outside of the sphere can not be disturbed by GLHUA sphere. By GL full wave method, we rigorously proved the incident EM wave field excited in outside of GLHUA sphere and propagation through the sphere without any scattering by the sphere, the total EM field in outside of the sphere equal to the incident wave field. Moreover, we prove that in GLHUA sphere with the pre cloak material, when r is going to origin, EM wave field propagation in GLHUA sphere is going to zero. \\
As new content of version 2 of this paper arxiv:1701.02583, We propose a novel $N$ dimensional Maxwell equation that is different from Gauge potential Maxwell field equations.
We propose a new $N$ dimensional Curl operator. The 3 dimensional curl operator is well defined, Howover, for $N$, $N \ne 3$ dimensinal space, the curl operator is never  be defined in other publised papers before. As a breakthough, we propose a new curl operator 
in any $N$ dimensinal space, $Curl_F  = (\nabla  \times )_N $:
\[
Curl_F  = (\nabla  \times )_N  = \left( {\nabla \nabla  \cdot  - \nabla  \cdot \nabla } \right)^{\frac{1}{2}} , (122)      
\]
and a $N$ Dimensional Maxwell's Equaions:
\[
\begin{array}{l}
 Curl_F E = (\nabla  \times )_N E =  - \frac{{\partial B}}{{\partial t}}, \\ 
 \nabla  \cdot B = 0, \\ 
 Curl_F H = (\nabla  \times )_N H = \frac{{\partial D}}{{\partial t}} + J, \\ 
 \nabla  \cdot D = \rho ,                    \\
 \end{array}   (123)
\]
For electric current or magnetic source, the analytic electromagnetic wave field solutions of the new $N$ dimensional Maxwell's equations (123) are obtained.

For $3$ dimensional equations, our curl operator $Curl_F $ in (122) become standard 3D $curl$ 
\[
curl_F E = (\nabla  \times )_3 E = \left| {\begin{array}{*{20}c}
   {\vec x_1 } & {\vec x_2 } & {\vec x_3 }  \\
   {\frac{\partial }{{\partial x_1 }}} & {\frac{\partial }{{\partial x_2 }}} & {\frac{\partial }{{\partial x_3 }}}  \\
   {E_{x_1 } } & {E_{x_2 } } & {E_{x_3 } }  \\
\end{array}} \right|    (124)                \\
\]

Our $N$ dimensional Maxwell equations in (123) becomes standard 3D Maxwell's equations.  
\[
\begin{array}{l}
 \nabla  \times E = \left| {\begin{array}{*{20}c}
   {\vec x_1 } & {\vec x_2 } & {\vec x_3 }  \\
   {\frac{\partial }{{\partial x_1 }}} & {\frac{\partial }{{\partial x_2 }}} & {\frac{\partial }{{\partial x_3 }}}  \\
   {E_{x_1 } } & {E_{x_2 } } & {E_{x_3 } }  \\
\end{array}} \right| =  - \frac{{\partial B}}{{\partial t}}, \\ 
 \nabla  \times H = \left| {\begin{array}{*{20}c}
   {\vec x_1 } & {\vec x_2 } & {\vec x_3 }  \\
   {\frac{\partial }{{\partial x_1 }}} & {\frac{\partial }{{\partial x_2 }}} & {\frac{\partial }{{\partial x_3 }}}  \\
   {H_{x_1 } } & {H_{x_2 } } & {H_{x_3 } }  \\
\end{array}} \right| = \frac{{\partial D}}{{\partial t}} + \vec J, \\ 
 \nabla  \cdot B = 0, \\ 
 \nabla  \cdot D = \rho , \\ 
 \end{array} (125) \\
\]

Our $N$ dimensional Maxwell's equations have great role in the theoretical and practicable applications,  in particular, in electromgnetic invisible cloaks and in super sciences. This breakthough discover has been announced in March 15, 2020 in Hunan Super Computational Society . The second author Dr. Feng Xie in GL Geophysical Laboratory and Stanford University made main and great contribution in this breakthough discovery and creation

All copyright and patent of the GLHUA EM cloaks,GLHUA sphere and GL modeling
and inversion
methods are reserved by authors in GL Geophysical Laboratory.
\end{abstract}

\maketitle

\section{\label{sec:level1}INTRODUCTION} 
Using GL no scattering modeling and inversion and many GL method simulation [1] [2] [3], we find a class nonzero relative anisotropic parameter solution of EM zero scattering inversion in the sphere $r \le R_2 $ We create a novel material in GLHUA sphere with relative EM parameter not less than 1 that makes the sphere is invisible. The parameters and their derivative are continuous across the boundary $r=R_2$ and the parameters are going to infinity at origin $r=0$. The phase velocity of EM wave in the sphere is less than light speed and going to zero at origin. We discovered and proved an essential property that in the local sphere $r \le R_O $  without EM source, including 
$R_2  < r < R_O $ annular layer in free space and  $r \le R_2 $ in GLHUA sphere, on any spherical surface with radius $r<R_O $, the spherical surface integral of $E_r \sin \theta $ and $H_r \sin \theta $ is zero, 
($E_r $ is radial electric wave field, $H_r $ is radial magnetic wave field). The essential property of EM wave is a key difference from the acoustic wave and seismic wave. Based on the essential property, by GL full wave method, we rigorously proved the incident EM wave field excited in the outside of GLHUA sphere and propagation through the sphere without any scattering from GLHUA sphere, the total EM field in the outside of GLHUA sphere equal to the incident wave field. EM wave field excited in the outside of the sphere can not be disturbed by GLHUA sphere. GLHUA sphere is complete invisible. Moreover, based on the above essential property of EM wave, we prove that in GLHUA sphere with the pre cloak material, when $r$ is going to origin, EM wave field propagation in GLHUA sphere is going to zero. We propose a special GL transform to map GLHUA outer annular layer cloak to GLHUA sphere. The anisotropic relative parameters of EM material in physical GLHUA outer annular layer cloak, $\varepsilon _{p,r}  = \mu _{p,r}  = 1$,
$\varepsilon _{p,\theta }  = \frac{1}{2}\left( {\frac{{r - R_1 }}{{R_2  - R_1 }} + \frac{{R_2  - R_1 }}{{r - R_1 }}} \right)$,
$\varepsilon _{p,\theta }  = \varepsilon _{p,\phi }  = \mu _{p,\theta }  = \mu _{p,\phi } $ are mapping to anisotropic relative parameter material in GLHUA invisible sphere that satisfy GLHUA pre invisible cloak material conditions. From the theoretical proof in GLHUA invisible sphere, we can rigorously prove GLHUA outer annular layer cloak is invisible cloak with concealment.  
The major new ingredients in this paper are proposed in 7 sections.The introduction is presented in the section 1. Second order Maxwell electromagnetic equation in anisotropic material in spherical coordinate, and the basic fundamental electromagnetic wave in free space is presented in the section 2. The Global and Local (GL) method for radial electromagnetic equation and the essential property of radial EM wave are presented in the section 3. In the section 4, we propose GL EM wave Greens equation and GL EM Greens function. GL EM Integral equation and its 
theoretical proof are proposed in the section 5. In the section 6, we propose GLHUA EM invisible sphere, in this section, we rigorously proved that EM wave field propagation is going to zero at origin in GLHUA sphere. In section 7, the discussion and conclusion is presented.

\section {Second Order Maxwell Electromagnetic Equation In Spherical Coordinate}
 
\subsection{Electromagnetic Maxwell Equation in Sphere Coordinate System}
\begin{equation}
\begin{array}{l}
 \nabla  \times \vec E = \frac{1}{{r^2 \sin \theta }}\left[ {\begin{array}{*{20}c}
   {\vec r} & {r\vec \theta } & {r\sin \theta \vec \phi }  \\
   {\frac{\partial }{{\partial r}}} & {\frac{\partial }{{\partial \theta }}} & {\frac{\partial }{{\partial \phi }}}  \\
   {E_r } & {rE_\theta  } & {r\sin \theta E_\phi  }  \\
\end{array}} \right] \\ 
  =  - i\omega \left( {H_r ,H_\theta  ,H_\phi  } \right)\left[ {\begin{array}{*{20}c}
   {\mu _r } & {} & {}  \\
   {} & {\mu _\theta  } & {}  \\
   {} & {} & {\mu _\phi  }  \\
\end{array}} \right]\mu _0  \\ 
 \end{array}
\end{equation}
\begin{equation}
\begin{array}{l}
 \nabla  \times \vec H = \frac{1}{{r^2 \sin \theta }}\left[ {\begin{array}{*{20}c}
   {\vec r} & {r\vec \theta } & {r\sin \theta \vec \phi }  \\
   {\frac{\partial }{{\partial r}}} & {\frac{\partial }{{\partial \theta }}} & {\frac{\partial }{{\partial \phi }}}  \\
   {H_r } & {rH_\theta  } & {r\sin \theta H_\phi  }  \\
\end{array}} \right] \\ 
  = i\omega \left( {E_r ,E_\theta  ,E_\phi  } \right)\left[ {\begin{array}{*{20}c}
   {\varepsilon _r } & {} & {}  \\
   {} & {\varepsilon _\theta  } & {}  \\
   {} & {} & {\varepsilon _\phi  }  \\
\end{array}} \right]\varepsilon _0  + \delta (\vec r - \vec r_s )\vec e_s  \\ 
 \end{array}
\end{equation}

\begin{equation}
\begin{array}{l}
 \nabla  \cdot \vec B = \frac{1}{{r^2 }}\frac{\partial }{{\partial r}}\left( {r^2 \mu _r H_r } \right) + \frac{1}{{\sin \theta r}}\frac{\partial }{{\partial \theta }}\sin \theta \mu _\theta  H_\theta   \\ 
  + \frac{1}{{r\sin \theta }}\frac{\partial }{{\partial \phi }}\mu _\phi  H_\phi   = 0, \\ 
 \end{array}
\end{equation}

\begin{equation}
\begin{array}{l}
 \nabla  \cdot \vec D = \frac{1}{{r^2 }}\frac{\partial }{{\partial r}}\left( {r^2 \varepsilon _r E_r } \right) + \frac{1}{{\sin \theta r}}\frac{\partial }{{\partial \theta }}\sin \theta \varepsilon _\theta  E_\theta   \\ 
  + \frac{1}{{r\sin \theta }}\frac{\partial }{{\partial \phi }}\varepsilon _\phi  E_\phi   = 4\pi \rho , \\ 
 \end{array}
\end{equation}
where $\rho $ is the electric charge, $\varepsilon _0 $ is the basic constant electric permittivity, $\mu _0 $ is the basic constant magnetic permeability, $\varepsilon _r $ , relative radial electric permittivity, $\varepsilon _\theta  $ and $\varepsilon _\phi  $, relative angular electric permittivity, $\mu _r $ , relative radial magnetic permeability, $\mu _\theta  $ and $\mu _\phi  $, relative angular magnetic permeability, $E_r$ is the radial electric field, $E_\theta  $ is the electric field in angular $\theta $ direction, $E_\phi  $ is the electric field in angular $\phi $ direction, $\vec E = \left( {E_r ,E_\theta  ,E_\phi  } \right)$ is the electric field vector, $\vec D$ is the electric displacement vector, $\vec B$ is the magnetic flux vector, $H_r$ is the radial magnetic field, $H_\theta  $  is the magnetic field in angular $\theta $ direction, $H_\phi  $  is the magnetic field in angular  $\phi $ direction,   $\vec H = \left( {H_r ,H_\theta  ,H_\phi  } \right)$  is the magnetic field vector, $\omega  = 2\pi f$ is the angular frequency, $f$ is the frequency $Hz$, $\vec r = (r,\theta ,\phi )$, the vector in spherical coordinate.
\subsection{Second Order Maxwell Electromagnetic Equation On Radial Electric Wave In Spherical Coordinate}
We use Electric-Magnetic-Electric EME translate process to translate Maxwell Equation in Spherical Coordinate (S1) and (S2) with anisotropic material into the Second order Maxwell Electromagnetic Equation on Radial Electric Wave Field In Spherical Coordinate

\begin{equation}
\begin{array}{l}
 \frac{\partial }{{\partial r}}\frac{1}{{\varepsilon _\theta  }}\frac{\partial }{{\partial r}}\varepsilon _r r^2 E_r  + \frac{1}{{\sin \theta }}\frac{\partial }{{\partial \theta }}\sin \theta \frac{{\partial E_r }}{{\partial \theta }} +  \\ 
 \frac{1}{{\sin ^2 \theta }}\frac{{\partial ^2 E_r }}{{\partial \phi ^2 }} + k^2 \mu _\theta  \varepsilon _r r^2 E_r  \\ 
  =  - \frac{1}{{\sin \theta }}\frac{\partial }{{\partial \theta }}\sin \theta \frac{\partial }{{\partial r}}\left( {\frac{1}{{i\omega \varepsilon _\theta  }}rJ_\theta  } \right) \\ 
  - \frac{1}{{\sin \theta }}\frac{\partial }{{\partial \phi }}\frac{\partial }{{\partial r}}\left( {\frac{1}{{i\omega \varepsilon _\theta  }}rJ_\phi  } \right) \\ 
  - \frac{\partial }{{\partial r}}\frac{1}{{i\omega \varepsilon _\theta  }}\frac{\partial }{{\partial r}}r^2 J_r  - ( - i\omega \mu _\theta  )r^2 J_r , \\ 
 \end{array}
\end{equation}
\subsection{Second order Maxwell Electromagnetic Equation on Radial Magnetic Wave Field In Spherical Coordinate}
We use MEM Magnetic-Electric-Magnetic translate process to translate Maxwell equation in spherical Coordinate (S1) and (S2) into the Second order Maxwell magnetic equation on radial magnetic wave field in spherical coordinate with anisotropic material,

\begin{equation}
\begin{array}{l}
 \frac{\partial }{{\partial r}}\frac{1}{{\left( {\mu _\theta  } \right)}}\frac{{\partial \left( {\mu _r r^2 H_r } \right)}}{{\partial r}} + \frac{1}{{\sin \theta }}\frac{\partial }{{\partial \theta }}\sin \theta \frac{{\partial H_r }}{{\partial \theta }} \\ 
  + \frac{1}{{\sin ^2 \theta }}\frac{{\partial H_r }}{{\partial \phi ^2 }} + k^2 \varepsilon _\theta  \mu _r r^2 H_r  \\ 
  = \frac{1}{{\sin \theta }}\frac{\partial }{{\partial \theta }}\sin \theta rJ_\phi   - \frac{1}{{\sin \theta }}\frac{\partial }{{\partial \phi }}rJ_\theta  , \\ 
 \end{array}
\end{equation}

\subsection{Second Order Maxwell Electromagnetic Equation On Angular Electric Wave Field In Spherical Coordinate}

After solving equation (5) and (6) and obtaining the radial electric wave field $E_r (r,\theta ,\phi )$ and radial magnetic wave field $H_r (r,\theta ,\phi )$, the following second order Maxwell electromagnetic equation govern angular electric wave field $E_\theta  (r,\theta ,\phi )$ and $E_\phi  (r,\theta ,\phi )$
\begin{equation}
\begin{array}{l}
 \frac{\partial }{{\partial r}}\frac{1}{{\left( {\mu _\theta  } \right)}}\frac{{\partial rE_\theta  }}{{\partial r}} + k^2 \varepsilon _\theta  rE_\theta   \\ 
  =  - i\omega \mu _0 \frac{1}{{\sin \theta }}\frac{{\partial H_r }}{{\partial \phi }} + \frac{\partial }{{\partial r}}\left( {\frac{1}{{\mu _\theta  }}\frac{{\partial E_r }}{{\partial \theta }}} \right) \\ 
  - i\omega \mu _0 rJ_\theta   \\ 
 \end{array}
\end{equation}

\begin{equation}
\begin{array}{l}
 \frac{\partial }{{\partial r}}\frac{1}{{\left( {\mu _\theta  } \right)}}\frac{{\partial rE_\phi  }}{{\partial r}} + k^2 \varepsilon _\theta  rE_\phi   \\ 
  = \frac{\partial }{{\partial r}}\frac{1}{{\left( {\mu _\theta  } \right)}}\frac{1}{{\sin \theta }}\frac{{\partial E_r }}{{\partial \phi }} + i\omega \mu _0 \frac{{\partial H_r }}{{\partial \theta }} \\ 
  - i\omega \mu _0 rJ_\phi   \\ 
 \end{array}
\end{equation}

\subsection{Second Order Maxwell Electromagnetic Equation on Angular Magnetic Wave Field In Spherical Coordinate}

After solving equation (5) and (6) and obtaining the radial electric wave field $E_r (r,\theta ,\phi )$  and radial magnetic wave field $H_r (r,\theta ,\phi )$ , the following second order Maxwell electromagnetic equation govern angular magnetic wave field $H_\theta  (r,\theta ,\phi )$ and $H_\phi  (r,\theta ,\phi )$

\begin{equation}
\begin{array}{l}
 \frac{\partial }{{\partial r}}\left( {\frac{1}{{\varepsilon _\theta  }}\frac{\partial }{{\partial r}}rH_\theta  } \right) + k^2 \mu _\theta  rH_\theta   =  \\ 
 i\omega \varepsilon _0 \frac{1}{{\sin \theta }}\frac{{\partial E_r }}{{\partial \phi }} + \frac{\partial }{{\partial r}}\left( {\frac{1}{{\varepsilon _\theta  }}\left( {\frac{{\partial H_r }}{{\partial \theta }}} \right)} \right) \\ 
  + \frac{\partial }{{\partial r}}\left( {\frac{1}{{\varepsilon _\theta  }}rJ_\phi  } \right), \\ 
 \end{array}
\end{equation}

\begin{equation}
\begin{array}{l}
 \frac{\partial }{{\partial r}}\left( {\frac{1}{{\varepsilon _\theta  }}\frac{\partial }{{\partial r}}rH_\phi  } \right) + k^2 \mu _\theta  rH_\phi   =  \\ 
 \frac{1}{{\sin \theta }}\frac{\partial }{{\partial r}}\frac{1}{{\varepsilon _\theta  }}\frac{{\partial H_r }}{{\partial \phi }} - i\omega \varepsilon _0 \frac{{\partial E_r }}{{\partial \theta }} \\ 
  - \frac{\partial }{{\partial r}}\left( {\frac{1}{{\varepsilon _\theta  }}rJ_\theta  } \right), \\ 
 \end{array}
\end{equation}

\subsection{Fundamental acoustic wave field}
Let $g(\vec r,\vec r_s )$ is the fundamental acoustic wave field, which is solution of following acoustic equation

\begin{equation}
\begin{array}{l}
 \frac{\partial }{{\partial r}}r^2 \frac{{\partial g}}{{\partial r}} + \frac{1}{{\sin \theta }}\frac{\partial }{{\partial \theta }}\sin \theta \frac{{\partial g}}{{\partial \theta }} + \frac{1}{{\sin ^2 \theta }}\frac{{\partial ^2 g}}{{\partial \phi ^2 }} \\ 
  + k^2 r^2 g = \delta (r - r_s )\frac{{\delta (\theta  - \theta _s )}}{{\sin \theta }}\delta (\phi  - \phi _s ), \\ 
 \end{array}
\end{equation}

\begin{equation}
g(\vec r,\vec r_s ) =  - \frac{1}{{4\pi }}\frac{{e^{ - ik\left| {\vec r - \vec r_s } \right|} }}{{\left| {\vec r - \vec r_s } \right|}}
\end{equation}

\begin{equation}
\begin{array}{l}
 \left| {\vec r - \vec r_s } \right| = \sqrt {\left| {\vec r - \vec r_s } \right|^2 } , \\ 
 \left| {\vec r - \vec r_s } \right|^2  = r^2  + r_s ^2  \\ 
  - 2rr_s \sin \theta \sin \theta _s \cos (\phi  - \phi _s ) \\ 
  - 2rr_s \cos \theta \cos \theta _s , \\ 
 \end{array}
\end{equation}
where $\vec r{}_s = (r_s ,\theta _s ,\phi _s )$ is point source location, 
$\vec r = (r,\theta ,\phi )$
 is variable spherical coordinate, i.e. observation point.

\subsection{Basic fundamental electromagnetic wave field in free space}
The basic fundamental electromagnetic wave field in free space can be excited by current source $\vec J = \delta (\vec r - \vec r{}_s)\vec e{}_j$, or magnetic moment source $\vec M = \delta (\vec r - \vec r{}_s)\vec e{}_j$.We consider the current source in this paper, similarly theorem and proof are suitable for the magnetic moment source. Let $\varepsilon _r  = \varepsilon _\theta   = \varepsilon _\phi   = 1$ and $\mu _r  = \mu _\theta   = \mu _\phi   = 1$, electric current source $\vec J = \delta (\vec r - \vec r{}_s)\vec e{}_j$,The basic fundamental electromagnetic wave field in free space are solutions of equations (5)-(6) in free space,

\begin{equation}
E^b {}_j = \frac{1}{{i\omega \varepsilon _0 }}\left( {\nabla \nabla  \cdot (g\vec e_j ) + k^2 g\vec e_j } \right)
\end{equation}

\begin{equation}
\begin{array}{l}
 E^b {}_{j,r} = \frac{1}{{i\omega \varepsilon _0 }}\left( {e_{jr} \left( {\frac{{\partial ^2 g}}{{\partial r^2 }} + k^2 g} \right) + e_{j\theta } \frac{\partial }{{\partial r}}\frac{1}{r}\frac{{\partial g}}{{\partial \theta }}} \right) \\ 
  + e_{j\phi } \frac{1}{{i\omega \varepsilon _0 }}\frac{\partial }{{\partial r}}\frac{1}{{r\sin \theta }}\frac{{\partial g}}{{\partial \phi }} \\ 
 \end{array}
\end{equation}
\begin{equation}
\begin{array}{l}
 H^b {}_j = \nabla  \times (g\vec e_j ), \\ 
 H^b {}_{j,r} = \left( {e_{j\phi } \frac{1}{r}\frac{{\partial g}}{{\partial \theta }} - e_{j\theta } \frac{1}{{r\sin \theta }}\frac{{\partial g}}{{\partial \phi }}} \right), \\ 
 \end{array}
\end{equation}
where the above equation are in the spherical coordinate system, $g(\vec r,\vec r_s )$ is denoted in (11)-(13), $\nabla  \cdot $ is diverge operator $E_j^b $  is basic fundamental electric wave, , $H_j^b $  is basic fundamental magnetic wave, which is excited by source $\vec J = \delta (\vec r - \vec r{}_s)\vec e{}_j$,  the up script $b $ means basic fundamental electric wave, the lower script $j$ means source  with excited vector $\vec e{}_j$. $E_{j,r}^b $ is radial basic fundamental electric wave, i.e. r component of $E_j^b $  , $H_{j,r}^b$  is radial basic fundamental magnetic wave, i.e. r component of $H_j^b $  , $\vec e{}_j$, j=1,2,3, is a unite vector,

\begin{equation}
\begin{array}{l}
 e_1  = e_x  = \left( {\begin{array}{*{20}c}
   {\sin \theta \cos \phi } & {\cos \theta \cos \phi } & { - \sin \phi }  \\
\end{array}} \right), \\ 
 e_2  = e_y  = \left( {\begin{array}{*{20}c}
   {\sin \theta \sin \phi } & {\cos \theta \sin \phi } & {\cos \phi }  \\
\end{array}} \right), \\ 
 e_3  = e_z  = \left( {\begin{array}{*{20}c}
   {\cos \theta } & { - \sin \theta } & 0  \\
\end{array}} \right), \\ 
 \end{array}
\end{equation}
\section{Global and Local (GL) method for radial electromagnetic equation and the essential property  of the radial electromagnetic wave}
In the section 1, we proposed second order Maxwell electromagnetic equation in anisotropic material in spherical coordinate. The fundamental electromagnetic wave field in free space in sphere coordinate is presented in section 2. These basic equations are used in this and next sections.

\subsection{Global and Local GL radial electric second order differential equation}
For $R_2  > 0$, we consider electromagnetic equation (5)-(10) in the sphere $r \le R_2$ with anisotropic media, in the outside of the sphere, $r > R_2 $ with basic isotropic electric permittivity $\varepsilon _0 $ and magnetic permeability $\mu _0 $ in free space. 
In our paper, we suppose that the electromagnetic source set is bounded, the bounded source set $\Omega _s $ is in outside of the large sphere with radius $R_O $, $R_O  > R_2 $,for example, the point source located in outside of the sphere, $r_s  > R_O  > R_2 $. 
 
Using Global and Local (GL) field method, we propose radial GL electromagnetic field and GL electromagnetic equation. and study the following GL radial electromagnetic equation in the sphere$r \le R_2 $.

Definition of GL radial electromagnetic wave field

\begin{equation}
\begin{array}{l}
 E(\vec r) = \varepsilon _r r^2 E_r (\vec r), \\ 
 H(\vec r) = \mu _r r^2 H_r (\vec r), \\ 
 \end{array}
\end{equation}
From the radial electric equation (5) , we propose GL radial electric second order differential equation
\begin{equation}
\begin{array}{l}
 \frac{\partial }{{\partial r}}\frac{1}{{\varepsilon _\theta  }}\frac{\partial }{{\partial r}}E + \frac{1}{{\varepsilon _r r^2 }}\frac{1}{{\sin \theta }}\frac{\partial }{{\partial \theta }}\sin \theta \frac{{\partial E}}{{\partial \theta }} \\ 
  + \frac{1}{{\varepsilon _r r^2 }}\frac{1}{{\sin ^2 \theta }}\frac{{\partial ^2 E}}{{\partial \phi ^2 }} + k^2 \mu _\theta  E = J_s , \\ 
 \end{array}
\end{equation}

where $k = 2\pi f\sqrt {\varepsilon _0 \mu _0 } $, 
\[
\begin{array}{l}
 J_s  =  \\ 
 \delta (r - r_s )\frac{1}{{\sin \theta }}\delta (\theta  - \theta _s )\delta (\phi  - \phi _s )e_r  \\ 
 \end{array}
\]
electric point source $r_s  > R_O $. The incident GL electric wave field in free space,

\begin{equation}
\begin{array}{l}
 E_j^b (\vec r) = r^2 E_{j,r}^b  = \frac{1}{{i\omega \varepsilon _0 }}r^2 e_{jr} \left( {\frac{{\partial ^2 g}}{{\partial r^2 }} + k^2 g} \right) \\ 
  + \frac{1}{{i\omega \varepsilon _0 }}r^2 \left( {e_{j\theta } \frac{\partial }{{\partial r}}\frac{1}{r}\frac{{\partial g}}{{\partial \theta }} + e_{j\phi } \frac{\partial }{{\partial r}}\frac{1}{{r\sin \theta }}\frac{{\partial g}}{{\partial \phi }}} \right), \\ 
 \end{array}
\end{equation}
where $g = g(r,r_s )$ is the fundamental solution of the acoustic wave equation, (11-13).
Let $E^b (\vec r)$ to denote one of $E_j^b (\vec r) = r^2 E_{j,r}^b $, $j = 1,2,3,$ it obvious that
\begin{equation}
\mathop {\lim }\limits_{r \to 0} E^b (\vec r) = 0,
\end{equation}
\begin{equation}
\mathop {\lim }\limits_{r \to 0} \frac{\partial }{{\partial r}}E^b (\vec r) = 0,
\end{equation}

\subsection{Global and Local GL radial magnetic field second order differential equation}
By definition of GL magnetic wave $H(\vec r) = \mu _r r^2 H_r (\vec r)$ in (18), from the radial magnetic equation (6), we propose GL radial magnetic field second order differential equation

\begin{equation}
\begin{array}{l}
 \frac{\partial }{{\partial r}}\frac{1}{{\mu _\theta  }}\frac{\partial }{{\partial r}}H + \frac{1}{{\mu _r r^2 }}\frac{1}{{\sin \theta }}\frac{\partial }{{\partial \theta }}\sin \theta \frac{{\partial H}}{{\partial \theta }} \\ 
  + \frac{1}{{\mu _r r^2 }}\frac{1}{{\sin ^2 \theta }}\frac{{\partial ^2 H}}{{\partial \phi ^2 }} + k^2 \varepsilon _\theta  H = M_s , \\ 
 \end{array}
\end{equation}
$M_s $ is magnetic point source, incident GL magnetic wave in free space is,
\begin{equation}
\begin{array}{l}
 H_j^b  = r^2 H_{j,r}^b  \\ 
  = r\left( {e_{j\phi } \frac{\partial }{{\partial \theta }}g - e_{j\theta } \frac{1}{{\sin \theta }}\frac{\partial }{{\partial \phi }}g} \right), \\ 
 \end{array}
\end{equation}
Let $H^b (\vec r)$ to denote one of the $H_j^b (\vec r) = r^2 H_{j,r}^b $, $j = 1,2,3,$ it obvious
\begin{equation}
\mathop {\lim }\limits_{r \to 0} H^b (\vec r) = 0,
\end{equation}
\begin{equation}
\mathop {\lim }\limits_{r \to 0} \frac{\partial }{{\partial r}}H^b (\vec r) = 0,
\end{equation}

\subsection{Spherical surface integral of incident GL electromagnetic wave is vanished
}
Define spherical surface integral of incident GL electric wave in free space as

\begin{equation}
E_0^b (r) = \frac{1}{{4\pi }}\int_0^\pi  {\int_0^{2\pi } {E^b (\vec r)\sin \theta d\theta d\phi } } ,
\end{equation}
Define spherical surface integral of incident GL magnetic wave in free space as

\begin{equation}
H_0^b (r) = \frac{1}{{4\pi }}\int_0^\pi  {\int_0^{2\pi } {H^b (\vec r)\sin \theta d\theta d\phi } } ,
\end{equation}
\subsection{Essential property of GL radial electromagnetic wave }
 ${\boldsymbol{Theorem \ 3.1: }}$,\ Suppose that the electromagnetic source set is bounded, the bounded source set is in outside of the sphere with large radius $R_O$, $r_s > R_O$. In the no source domain, with weight $\sin \theta$ spherical surface integral of incident radial GL electromagnetic wave is zero.

\begin{equation}
E_0^b (r) = 0,H_0^b (r) = 0,
\end{equation}
The spherical surface integral of radial GL electromagnetic wave is zero,

\begin{equation}
E_0 (r) = 0,H_0 (r) = 0,
\end{equation}
 ${\boldsymbol{Proof : }}$,\ By equation (19), the GL radial electric second order differential equation in free space is  

\begin{equation}
\begin{array}{l}
 \frac{{\partial ^2 }}{{\partial r^2 }}E^b  + \frac{1}{{r^2 }}\frac{1}{{\sin \theta }}\frac{\partial }{{\partial \theta }}\sin \theta \frac{{\partial E^b }}{{\partial \theta }} \\ 
  + \frac{1}{{r^2 }}\frac{1}{{\sin ^2 \theta }}\frac{{\partial ^2 E^b }}{{\partial \phi ^2 }} + k^2 E^b  = S_E  \\ 
 \end{array}
\end{equation}
The electromagnetic source $S_E$ is denoted by (5), (6), because the bounded source set is in outside of the sphere, with large radius $R_O$,$r_s > R_O$, there exist the no source domain $r < R_O$ ,  in the no source domain or for plane electromagnetic wave without source,$S_E = 0$. Use $\sin \theta$ times both sides of (31) and take spherical surface integral and by integral by parts, we get Linville ordinary equation

\begin{equation}
\frac{{d^2 }}{{dr^2 }}E_0^b  + k^2 E_0^b  = 0,
\end{equation}
From (21) and (22), the initial condition is
\begin{equation}
\mathop {\lim }\limits_{r \to 0} E_0^b (r) = 0
\end{equation}
\begin{equation}
\mathop {\lim }\limits_{r \to 0} \frac{\partial }{{\partial r}}E_0^b ( r) = 0,
\end{equation}
The equation system (32)-(34) has only zero solution. We have proved that 
$E_0 ^b (r)$ is complete vanished 
\[
E_0^b (r) = 0,
\]
Similarly, we have proved that $H_0 ^b (r)$ is complete vanished,
\[
H_0^b (r) = 0.
\]
The first part (29) of the theorem 3.1 is proved.  Next, we prove the second part of the theorem. Suppose that the electromagnetic source set is bounded, the bounded source set is in outside of the sphere with radius $R_O$, in the GLHUA sphere, $r \le R_2  < R_O $, the source term is zero in the right hand of GL electric equation (19). Use  $\sin \theta$  times both sides of (19) and take sphere surface integral and by integral by parts, we have ordinary equation

\begin{equation}
\begin{array}{l}
 \frac{\partial }{{\partial r}}\frac{1}{{\varepsilon _\theta  }}\frac{\partial }{{\partial r}}E_0 (r) + k^2 \mu _\theta  E_0 (r) = 0, \\ 
 r \le R_2 , \\ 
 \end{array}
\end{equation}
By above proof, we have initial condition $E_0^b (R_2 ) = 0$ and 
$\frac{\partial }{{\partial r}}E_0^b (R_2 ) = 0$. Because in the GLHUA sphere, electromagnetic material parameters and their derivative are continuous across the outer boundary $r=R_2$, the radial GL electromagnetic wave and their derivative are continuous across the boundary $r=R_2$.

\begin{equation}
E_0 (R_2 ) = E_0^b (R_2 ) = 0,
\end{equation}

\begin{equation}
\frac{\partial }{{\partial r}}E_0 (R_2 ) = \frac{\partial }{{\partial r}}E_0^b (R_2 ) = 0,
\end{equation}

The solution of ordinary differential equation (35) with zero initial boundary condition (36) and (37) must be zero. Therefore,$E_0 (r) = 0$ similar 
$H_0 (r) = 0$, we proved theorem 3.1 that spherical surface integral of GL electromagnetic wave is vanished. 
\hfill\break

${\boldsymbol{Theorem \ 3.2: }}$ \ Spherical surface integral of incident radial electromagnetic wave is zero. Spherical surface integral of radial electromagnetic wave is zero.
\hfill\break

 ${\boldsymbol{Proof : }}$,\ 
\begin{equation}
\begin{array}{l}
 {E_0 ^b} _r (r) =  \\ 
 \frac{1}{{4\pi r^2 }}\int_0^\pi  {\int_0^{2\pi } {E^b (\vec r)\sin \theta d\theta d\phi  = 0} } , \\ 
 \end{array}
\end{equation}

\begin{equation}
\begin{array}{l}
 {H_0 ^b} _r (r) =  \\ 
  = \frac{1}{{4\pi r^2 }}\int_0^\pi  {\int_0^{2\pi } {H^b (\vec r)\sin \theta d\theta d\phi  = 0} } , \\ 
 \end{array}
\end{equation}
\begin{equation}
\begin{array}{l}
 {E_0 }_{,r} (r) =  \\ 
  = \frac{1}{{4\pi r^2 }}\int_0^\pi  {\int_0^{2\pi } {E(\vec r)\sin \theta d\theta d\phi  = 0} } , \\ 
 \end{array}
\end{equation}

\begin{equation}
\begin{array}{l}
 {H_0 }_{,r} (r) =  \\ 
  = \frac{1}{{4\pi r^2 }}\int_0^\pi  {\int_0^{2\pi } {H(\vec r)\sin \theta d\theta d\phi  = 0} }  \\ 
 \end{array}
\end{equation}
For EM point source and $r_s > R_O$ , and $r< R_O$ ,also from (15) and (16), by direct integral, we can calculate

\begin{equation}
\begin{array}{l}
 {E_0 ^b} _{jr} (r) =  \\ 
  = \frac{1}{{4\pi }}\int_0^\pi  {\int_0^{2\pi } {E^b _{jr} (\vec r)\sin \theta d\theta d\phi  = 0,} }  \\ 
 \end{array}
\end{equation}
and 

\begin{equation}
\begin{array}{l}
 {H_0 ^b }_{jr} (r) =  \\ 
  = \frac{1}{{4\pi }}\int_0^\pi  {\int_0^{2\pi } {H^b _{jr} (\vec r)\sin \theta d\theta d\phi  = 0} ,}  \\ 
 \end{array}
\end{equation}
Here, we direct integral of (43), because (16)

\[
\begin{array}{l}
 H_{jr}^b  = \frac{1}{{r^2 \sin \theta }}\frac{\partial }{{\partial \theta }}(r\sin \theta e_{j\phi } g) \\ 
  - \frac{1}{{r^2 \sin \theta }}\frac{\partial }{{\partial \phi }}(re_{j\theta } g),j = 1,2,3, \\ 
 \end{array}
\]
$\vec e{}_j$ is an unit vector of the source, j=1,2,3, in (17), 

\[
\begin{array}{l}
 H_{0jr}^b (r) = \frac{1}{{4\pi }}\int_0^\pi  {\int_0^{2\pi } {H_{jr}^b (\vec r)\sin \theta d\theta d\phi } }  \\ 
  = \frac{1}{{4\pi }}\frac{1}{{r^2 }}\int_0^{2\pi } {\int_0^\pi  {\frac{\partial }{{\partial \theta }}(r\sin \theta e_{j\phi } g)d\theta d\phi } }  \\ 
  - \frac{1}{{4\pi }}\frac{1}{{r^2 }}\int_0^\pi  {\int_0^{2\pi } {\frac{\partial }{{\partial \phi }}(re_{j\theta } g)d\phi d\theta  = 0} } , \\ 
 \end{array}
\]
(43) is already proved by direct integral. Similarly, by direct integral, we can prove (42).

From theorem 3.1,$E_0 (r) = 0$, by definition of GL electromagnetic wave field

\[
\begin{array}{l}
 E(\vec r) = \varepsilon _r r^2 E_r (\vec r), \\ 
 H(\vec r) = \mu _r r^2 H_r (\vec r), \\ 
 \end{array}
\]
we have 
\[E_{0,r} (\vec r) = \frac{1}{{\varepsilon _r r^2 }}E_0 (\vec r) = 0\] and
 \[H_{0.r} (\vec r) = \frac{1}{{\mu _r r^2 }}H_0 (\vec r) = 0\]. The theorem 3.2 is proved.
\hfill\break

The spherical surface integral of incident radial electromagnetic wave is zero that is essential property. The key property is essential different between electromagnetic wave and acoustic wave. Note that GL electromagnetic wave (18) is not Maxwell electromagnetic field wave, also is not flux nor displace current. In the GL method, Global and Local virtual wave $E$ and $H$ in (18) is convenient under any coordinate transform. Global and Local virtual wave (18) and GL second order differential equation (19) (23) are important for study GLHUA sphere and GLHUA cloak.  It is shown that the GL electromagnetic differential equation and their incident wave (18)-(21) and (22)-(25) have sane equation form and theoretical properties. For simply, we study GL electric differential equation and its incident wave (18)-(21) in detail.

\section{ GL electromagnetic Greens equation}
\subsection{We propose GL electromagnetic Greens equation for GL electric equation (18) and GL magnetic equation (23)}
\begin{equation}
\begin{array}{l}
 \frac{\partial }{{\partial r}}\frac{\partial }{{\partial r}}G(\vec r,\vec r') + \frac{1}{{r^2 }}\frac{1}{{\sin \theta }}\frac{\partial }{{\partial \theta }}\sin \theta \frac{\partial }{{\partial \theta }}G(\vec r,\vec r') \\ 
  + \frac{1}{{r^2 }}\frac{1}{{\sin ^2 \theta }}\frac{{\partial ^2 }}{{\partial \phi ^2 }}G(\vec r,\vec r') \\ 
  + k^2 G(\vec r,\vec r') = \delta (\vec r - \vec r'), \\ 
 \end{array}
\end{equation}
\subsection{	GL electromagnetic Greens function}
We find and propose GL Greens function $G(\vec r,\vec r')$ which is the solution of above GL electromagnetic Greens equation (44),
\begin{equation}
G(\vec r,\vec r') = rr'g(\vec r,\vec r'),
\end{equation}
\begin{equation}
g(\vec r,\vec r') =  - \frac{1}{{4\pi }}\frac{{e^{ - ik\left| {\vec r - \vec r'} \right|} }}{{\left| {\vec r - \vec r'} \right|}},
\end{equation}
Where 
\[
\begin{array}{l}
 \left| {\vec r - \vec r'} \right| = \sqrt {\left| {\vec r - \vec r'} \right|^2 } , \\ 
 \left| {\vec r - \vec r'} \right|^2  = r^2  - 2rr'\sin \theta \sin \theta '\cos (\phi  - \phi ') \\ 
  - 2rr'\cos \theta \cos \theta ' + r'^2 , \\ 
 \end{array}
\]
 The GL Greens equation (44) and GL Greens function (45) are suitable for all global free space which is as background of local cloak. 

\subsection{The spherical surface integral of the GL electromagnetic Greens equation }
We take
\begin{equation}
\begin{array}{l}
 G_0 (r,r') =  \\ 
 \frac{1}{{4\pi }}\int_0^\pi  {\int_0^{2\pi } {G(\vec r,\vec r')\sin \theta d\theta d\phi } } , \\ 
 \end{array}
\end{equation}
Take sphere surface integral of GL electromagnetic Greens equation (44), then sphere surface integral of Greens function, ${G_0} (r,r')$, in (44) satisfy 
\begin{equation}
\begin{array}{l}
 \frac{\partial }{{\partial r}}\frac{\partial }{{\partial r}}G_0 (r,r') \\ 
  + k^2 G_0 (r,r') = \delta (r - r'), \\ 
 \end{array}
\end{equation}
We find spherical surface integral of GL Greens function $G_0 (r,r')$ which is the solution of above equation (48),
\begin{equation}
\begin{array}{l}
 G_0 (r,r') =  \\ 
 rr'ikj_0 (kr)\left( {j_0 (kr') - iy_0 (kr')} \right), \\ 
 r \le r', \\ 
 G_0 (r,r') =  \\ 
 rr'ikj_0 (kr')\left( {j_0 (kr) - iy_0 (kr)} \right), \\ 
 r' \le r, \\ 
 \end{array}
\end{equation}
\begin{equation}
\begin{array}{l}
 j_0 (kr) = \frac{{\sin kr}}{{kr}}, \\ 
 y_0 (kr) =  - \frac{{\cos kr}}{{kr}}, \\ 
 \end{array}
\end{equation}

\section{ GL Electromagnetic Integral Equation}

We propose the Global and Local GL integral equation on GL electric wave field $E(\vec r)$ in (18) in the sphere body $r \le R_2 $
\begin{equation}
\begin{array}{l}
 E(\vec r') = E^b (\vec r') -  \\ 
 \int\limits_{S(r \le R_2 )} {\left( {1 - \frac{1}{{\varepsilon _\theta  }}} \right)\frac{\partial }{{\partial r}}G\frac{\partial }{{\partial r}}EdV}  \\ 
  + \int\limits_{S(r \le R_2 )} {\frac{1}{{r^2 }}\left( {1 - \frac{1}{{\varepsilon _r }}} \right)}  \\ 
 \left( {\frac{1}{{\sin \theta }}\frac{\partial }{{\partial \theta }}\sin \theta \frac{{\partial G}}{{\partial \theta }} + \frac{1}{{\sin ^2 \theta }}\frac{{\partial ^2 G}}{{\partial \phi ^2 }}} \right)EdV \\ 
  + \int\limits_{S(r \le R_2 )} {k^2 \left( {1 - \mu _\theta  } \right)GEdV} , \\ 
 \end{array}
\end{equation}
The equivalent between GL integral equation (51) and GL electric wave differential equation (19) is proved next theorem 5.1. In integral equation (51), we change $E$   to $H$, $\varepsilon _r $ to $\mu _r $, 
$\varepsilon _\theta  $ to $\mu _\theta  $,  we obtain GL magnetic integral equation.

\begin{equation}
\begin{array}{l}
 H(\vec r') = H^b (\vec r') \\ 
  - \int\limits_{S(r \le R_2 )} {\left( {1 - \frac{1}{{\mu _\theta  }}} \right)\frac{\partial }{{\partial r}}G\frac{\partial }{{\partial r}}HdV}  \\ 
  + \int\limits_{S(r \le R_2 )} {\frac{1}{{r^2 }}\left( {1 - \frac{1}{{\mu _r }}} \right)}  \\ 
 \left( {\frac{1}{{\sin \theta }}\frac{\partial }{{\partial \theta }}\sin \theta \frac{{\partial G}}{{\partial \theta }} + \frac{1}{{\sin ^2 \theta }}\frac{{\partial ^2 G}}{{\partial \phi ^2 }}} \right)HdV \\ 
  + \int\limits_{S(r \le R_2 )} {k^2 \left( {1 - \varepsilon _\theta  } \right)GHdV} , \\ 
 \end{array}
\end{equation}

\subsection{ The equivalent between GL integral equation (51) and GL electric wave differential equation (19)}

${\boldsymbol{Theorem \ 5.1 : }}$ \ Suppose that the radial electric wave
 $E(\vec r)$ is solution of the GL radial electric wave differential equation (19) with incident wave (20)-(22), Greens function $G(\vec r,\vec r')$ in (45) satisfy the GL electromagnetic Greens differential equation (44), then 
$E(\vec r)$  satisfy the GL integral equation (51)
\hfill\break

${\boldsymbol{Proof: }}$ \ By using Greens function $G(\vec r,\vec r')$ in   
 $(45)$ to time the GL electric differential equation (19) and after doing some calculation and integral by part, we have
\begin{equation}
\begin{array}{l}
 \int\limits_{S(r \le R_2 )} {} \frac{\partial }{{\partial r}}\left( {\frac{1}{{\varepsilon _\theta  }}\left( {\frac{\partial }{{\partial r}}E} \right)G} \right)dV \\ 
  - \int\limits_{S(r \le R_2 )} {} \frac{1}{{\varepsilon _\theta  }}\frac{\partial }{{\partial r}}E\frac{\partial }{{\partial r}}GdV \\ 
  + \int\limits_{S(r \le R_2 )} {} \left( {\frac{1}{{\varepsilon _r r^2 }}} \right. \\ 
 \left. {\left( {\frac{1}{{\sin \theta }}\frac{\partial }{{\partial \theta }}\sin \theta \frac{{\partial G}}{{\partial \theta }} + \frac{1}{{\sin ^2 \theta }}\frac{{\partial ^2 G}}{{\partial \phi ^2 }}} \right)} \right)EdV \\ 
  + \int\limits_{S(r \le R_2 )} {} k^2 \mu _\theta  GEdV = E^b (\vec r) \\ 
 \end{array}
\end{equation}

By using unknown wave function $E(\vec r,\vec r_s )$ to time the GL Greens equation (44) and after do some calculation and integral by part, we have
\begin{equation}
\begin{array}{l}
 \int\limits_{S(r \le R_2 )} {\frac{\partial }{{\partial r}}\left( {\left( {\frac{\partial }{{\partial r}}G} \right)E} \right)dV}  \\ 
  - \int\limits_{^{S(r \le R_2 )} } {\frac{\partial }{{\partial r}}G\frac{\partial }{{\partial r}}EdV}  \\ 
  + \int\limits_{S(r \le R_2 )} {\left( {\frac{1}{{r^2 }}} \right.}  \\ 
 \left. {\left( {\frac{1}{{\sin \theta }}\frac{\partial }{{\partial \theta }}\sin \theta \frac{{\partial G}}{{\partial \theta }} + \frac{1}{{\sin ^2 \theta }}\frac{{\partial ^2 G}}{{\partial \phi ^2 }}} \right)} \right)EdV \\ 
  + \int\limits_{R^3 } {k^2 GEdV = E(\vec r)} , \\ 
 \end{array}
\end{equation}

To subtract (53) from (54)

\begin{equation}
\begin{array}{l}
 E(\vec r') = E^b (\vec r') \\ 
  + \frac{{e^{ - ikr'} }}{{4\pi }}\int_0^\pi  {\int_0^{2\pi } {E(0,\theta ,\phi )\sin \theta d\theta d\phi } }  \\ 
  - \int\limits_{S(r \le R_2 )} {\left( {1 - \frac{1}{{\varepsilon _\theta  }}} \right)\frac{\partial }{{\partial r}}G\frac{\partial }{{\partial r}}EdV}  \\ 
  + \int\limits_{S(r \le R_2 )} {\left( {\frac{1}{{r^2 }}\left( {1 - \frac{1}{{\varepsilon _r }}} \right)} \right.}  \\ 
 \left. {\left( {\frac{1}{{\sin \theta }}\frac{\partial }{{\partial \theta }}\sin \theta \frac{{\partial G}}{{\partial \theta }} + \frac{1}{{\sin ^2 \theta }}\frac{{\partial ^2 G}}{{\partial \phi ^2 }}} \right)} \right)EdV \\ 
  + \int\limits_{S(r \le R_2 )} {k^2 \left( {1 - \mu _\theta  } \right)GEdV} . \\ 
 \end{array}
\end{equation}

Substitute (27) for sphere surface integral of electric wave into the (55), because the anisotropic inhomogeneous electromagnetic relative material parameter are variable in the sphere,$r \le R_2$  The outside of sphere, $r> R_2$  is free space with relative electric permittivity $diag(1,1,1)$ and magnetic permeability $diag(1,1,1)$ , because (30) 

\[
\frac{{e^{ - ikr'} }}{{4\pi }}\int_0^\pi  {\int_0^{2\pi } {E(0,\theta ,\phi )\sin \theta d\theta d\phi  = 0} } 
\]
£¬

the integral equation (55) will become

\[
\begin{array}{l}
 E(\vec r') = E^b (\vec r') \\ 
  - \int\limits_{S(r \le R_2 )} {\left( {1 - \frac{1}{{\varepsilon _\theta  }}} \right)\frac{\partial }{{\partial r}}G\frac{\partial }{{\partial r}}EdV}  \\ 
  + \int\limits_{S(r \le R_2 )} {\left( {\frac{1}{{r^2 }}\left( {1 - \frac{1}{{\varepsilon _r }}} \right)} \right.}  \\ 
 \left. {\left( {\frac{1}{{\sin \theta }}\frac{\partial }{{\partial \theta }}\sin \theta \frac{{\partial G}}{{\partial \theta }} + \frac{1}{{\sin ^2 \theta }}\frac{{\partial ^2 G}}{{\partial \phi ^2 }}} \right)} \right)EdV \\ 
  + \int\limits_{S(r \le R_2 )} {k^2 \left( {1 - \mu _\theta  } \right)GEdV} . \ \ \ \ (51) \\ 
 \end{array}
\]
The theorem 5.1 is proved.

\section{Theory of GLHUA electromagnetic invisible sphere and behavior of the electromagnetic wave field propagation}
\subsection{GLHUA pre cloak material conditions on relative electric permittivity and magnetic permeability for invisible sphere }
In this section, we propose GLHUA pre cloak material conditions of anisotropic relative electric permittivity and magnetic permeability for invisible virtual sphere $r \le R_2 $, which are of the following properties for invisible sphere $r \le R_2 $. The electric permittivity is the product of relative electric permittivity and $\varepsilon _0 $, the magnetic permeability is the product of relative magnetic permeability and $\mu _0 $.$\varepsilon _0 $ Is the basic constant electric permittivity, $\mu _0 $ is basic constant magnetic permeability. GLHUA pre cloak material conditions in invisible virtual sphere  $r \le R_2 $  are as follows:

\begin{equation}
\begin{array}{l}
(6.1)\mu _r \left( r \right) = \varepsilon _r (r), \\ 
 \mu _\theta  \left( r \right) = \mu _\phi  \left( r \right) = \varepsilon _\theta  \left( r \right) = \varepsilon _\phi  \left( r \right), \\ 
 {\rm in \ the \ sphere } \ r \le R_2 \ {\rm  are \ continuous} \\ 
 {\rm \ differentiable \ function \ of \ r,                                        } \\ 
 \end{array}
\end{equation}
\begin{equation}
\begin{array}{l}
 (6.2){\rm these \ parameter \ functions } \\ 
 {\rm and \ their \ derivative \ functions } \\ 
 {\rm are \ continuous \ across \ boundary} \\ 
 {\rm  }r = R_2 {\rm , } \\ 
 \end{array}
\end{equation}
\begin{equation}
\begin{array}{l}
 \mathop {(6.3)\lim }\limits_{r \to 0} r^2 \mu _r \left( r \right) =  \\ 
 \mathop {\lim }\limits_{r \to 0} r^2 \varepsilon _r (r) = \infty , \\ 
 \end{array}
\end{equation}
\begin{equation}
\begin{array}{l}
 (6.4)\mu _\theta  \left( r \right) = \mu _\phi  \left( r \right) \\ 
  = \varepsilon _\theta  \left( r \right) = \varepsilon _\phi  \left( r \right) = f(r)\frac{1}{{r^2 }}, \\ 
 {\rm and \ their \ derivative \ are } \\ 
 {\rm continuous \ across \ boundary } \\ 
 r = R_2 ,\mathop {\lim }\limits_{r \to 0} f(r) = \frac{{R_2 ^2 }}{2}, \\ 
 \mathop {\lim }\limits_{r \to 0} \frac{1}{r}f'(r) = 0, \\ 
 \end{array}
\end{equation}
\subsection{GL electromagnetic field   and   are approaching to zero at r=0 in invisible sphere}
${\boldsymbol {Theorem \ 6.1: }}$ \ Suppose that the anisotropic relative electric permittivity$\varepsilon _r \left( r \right)$,$\varepsilon _\theta  \left( r \right)$ $\varepsilon _\phi  \left( r \right)$ and magnetic permeability, $\mu _r \left( r \right)$,$\mu _\theta  \left( r \right)$ $\mu _\phi  \left( r \right)$ , satisfy the above GLHUA pre cloak material conditions in invisible sphere  $(6.1) \ to \ (6.4)$, also we suppose that 
\begin{equation}
\int\limits_{S(r \le R_2 )} {\left( {\left| {E(\vec r)} \right|^2  + \left| {H(\vec r)} \right|^2 } \right)dV} is finite, 
\end{equation}
then GL radial electromagnetic wave field is vanished at origin,$r = 0$,
\begin{equation}
\mathop {\lim }\limits_{r \to 0} E(\vec r) = 0,
\end{equation}
\begin{equation}
\mathop {\lim }\limits_{r \to 0} H(\vec r) = 0,
\end{equation}
\hfill\break

${\boldsymbol {Proof£º}}$
\begin{equation}
\begin{array}{l}
 \mathop {\lim }\limits_{r \to 0} r^2 E(\vec r) = 0, \\ 
 \mathop {\lim }\limits_{r \to 0} r^2 H(\vec r) = 0, \\ 
 \end{array}
\end{equation}
is derived from (60). the additional condition (60) is reasonable finite energy condition. From GL radial electric integral equation (51)), we have
\begin{equation}
\begin{array}{l}
 E(\vec r') = E^b (\vec r') -  \\ 
  - \int\limits_{S(r \le R_2 )} {\frac{\partial }{{\partial r}}\left( {\left( {\left( {1 - \frac{1}{{\varepsilon _\theta  }}} \right)\frac{\partial }{{\partial r}}G} \right)E} \right)dV}  \\ 
  + \int\limits_{S(r \le R_2 )} {\frac{\partial }{{\partial r}}\left( {\left( {1 - \frac{1}{{\varepsilon _\theta  }}} \right)\frac{\partial }{{\partial r}}G} \right)EdV}  \\ 
  + \int\limits_{S(r \le R_2 )} {\left( {\frac{1}{{r^2 }}} \right.} \left( {1 - \frac{1}{{\varepsilon _r }}} \right) \\ 
 \left. {\left( {\frac{1}{{\sin \theta }}\frac{\partial }{{\partial \theta }}\sin \theta \frac{{\partial G}}{{\partial \theta }} + \frac{1}{{\sin ^2 \theta }}\frac{{\partial ^2 G}}{{\partial \phi ^2 }}} \right)} \right)EdV \\ 
  + \int\limits_{S(r \le R_2 )} {k^2 } \left( {1 - \mu _\theta  } \right)GEdV, \\ 
 \end{array}
\end{equation}
The integral equation (51) is translated to

\begin{equation}
\begin{array}{l}
 \frac{1}{{\varepsilon _\theta  }}E(\vec r') = E^b (\vec r') -  \\ 
  - \int\limits_{S(r \le R_2 )} {\left( {\frac{\partial }{{\partial r}}\frac{1}{{\varepsilon _\theta  }}} \right)\frac{{\partial G}}{{\partial r}}EdV}  \\ 
  + \int\limits_{S(r \le R_2 )} {\left( {1 - \frac{1}{{\varepsilon _\theta  }}} \right)\left( {\frac{{\partial ^2 }}{{\partial r^2 }}G} \right)EdV}  \\ 
  + \int\limits_{S(r \le R_2 )} {\left( {\frac{1}{{r^2 }}\left( {1 - \frac{1}{{\varepsilon _r }}} \right)} \right.}  \\ 
 \left. {\left( {\frac{1}{{\sin \theta }}\frac{\partial }{{\partial \theta }}\sin \theta \frac{{\partial G}}{{\partial \theta }} + \frac{1}{{\sin ^2 \theta }}\frac{{\partial ^2 G}}{{\partial \phi ^2 }}} \right)} \right)EdV \\ 
  + \int\limits_{S(r \le R_2 )} {k^2 \left( {1 - \mu _\theta  } \right)} GEdV, \\ 
 \end{array}
\end{equation}
The above equation (65) becomes
\begin{equation}
\begin{array}{l}
 \frac{1}{{\varepsilon _\theta  }}E(\vec r') = E^b (\vec r') -  \\ 
  - \int\limits_{S(r \le R_2 )} {\left( {\frac{\partial }{{\partial r}}\frac{1}{{\varepsilon _\theta  }}} \right)\frac{{\partial G}}{{\partial r}}EdV}  \\ 
  - \int\limits_{S(r \le R_2 )} {\left( {\frac{1}{{r^2 }}\left( {1 - \frac{1}{{\varepsilon _\theta  }}} \right)} \right.}  \\ 
 \left. {\left( {\frac{1}{{\sin \theta }}\frac{\partial }{{\partial \theta }}\sin \theta \frac{{\partial G}}{{\partial \theta }} + \frac{1}{{\sin ^2 \theta }}\frac{{\partial ^2 G}}{{\partial \phi ^2 }}} \right)} \right)EdV \\ 
  - \int\limits_{S(r \le R_2 )} {k^2 \left( {1 - \frac{1}{{\varepsilon _\theta  }}} \right)} GEdV \\ 
  + \int\limits_{S(r \le R_2 )} {\left( {\frac{1}{{r^2 }}\left( {1 - \frac{1}{{\varepsilon _r }}} \right)} \right.}  \\ 
 \left. {\left( {\frac{1}{{\sin \theta }}\frac{\partial }{{\partial \theta }}\sin \theta \frac{{\partial G}}{{\partial \theta }} + \frac{1}{{\sin ^2 \theta }}\frac{{\partial ^2 G}}{{\partial \phi ^2 }}} \right)} \right)EdV \\ 
  + \int\limits_{S(r \le R_2 )} {k^2 \left( {1 - \mu _\theta  } \right)} GEdV. \\ 
 \end{array}
\end{equation}
The GL electric integral equation (51) becomes to

\begin{equation}
\begin{array}{l}
 \frac{1}{{\varepsilon _\theta  }}E(\vec r') = E^b (\vec r') -  \\ 
  - \int\limits_{S(r \le R_2 )} {\left( {\frac{\partial }{{\partial r}}\frac{1}{{\varepsilon _\theta  }}} \right)\frac{{\partial G}}{{\partial r}}EdV}  \\ 
  + \int\limits_{S(r \le R_2 )} {\left( {\frac{1}{{r^2 }}\left( {\frac{1}{{\varepsilon _\theta  }} - \frac{1}{{\varepsilon _r }}} \right)} \right.}  \\ 
 \left. {\left( {\frac{1}{{\sin \theta }}\frac{\partial }{{\partial \theta }}\sin \theta \frac{{\partial G}}{{\partial \theta }} + \frac{1}{{\sin ^2 \theta }}\frac{{\partial ^2 G}}{{\partial \phi ^2 }}} \right)} \right)EdV \\ 
  + \int\limits_{S(r \le R_2 )} {k^2 \left( {\frac{1}{{\varepsilon _\theta  }} - \mu _\theta  } \right)GEdV} , \\ 
 \end{array}
\end{equation}

Let
\begin{equation}
G_L (r,r') = G(\vec r,\vec r') - G_0 (r,r')
\end{equation}

to substitute (68) for $G(\vec r,\vec r')$ into (67) , and by using $E_0 (r) = 0$ in (30), the integral equation (67) becomes

\begin{equation}
\begin{array}{l}
 \frac{1}{{\varepsilon _\theta  }}E(\vec r') = E^b (\vec r') -  \\ 
  - \int\limits_{S(r \le R_2 )} {\left( {\frac{\partial }{{\partial r}}\frac{1}{{\varepsilon _\theta  }}} \right)\frac{{\partial G_L }}{{\partial r}}EdV}  \\ 
  + \int\limits_{S(r \le R_2 )} {\left( {\frac{1}{{r^2 }}\left( {\frac{1}{{\varepsilon _\theta  }} - \frac{1}{{\varepsilon _r }}} \right)} \right.}  \\ 
 \left. {\left( {\frac{1}{{\sin \theta }}\frac{\partial }{{\partial \theta }}\sin \theta \frac{{\partial G_L }}{{\partial \theta }} + \frac{1}{{\sin ^2 \theta }}\frac{{\partial ^2 G_L }}{{\partial \phi ^2 }}} \right)} \right)EdV \\ 
  + \int\limits_{S(r \le R_2 )} {k^2 \left( {\frac{1}{{\varepsilon _\theta  }} - \mu _\theta  } \right)G_L EdV} , \\ 
 \end{array}
\end{equation}

where
\begin{equation}
\begin{array}{l}
 G_L (r,r') = G(\vec r,\vec r') - G_0 (r,r') \\ 
  = rr'\sum\limits_{l = 1}^\infty  {} g{}_l(r,r')\sum\limits_{m =  - l}^l {} Y_l^{m*} (\theta ,\phi )Y_l^m (\theta ',\phi '), \\ 
 \end{array}
\end{equation}
\begin{equation}
\begin{array}{l}
 g{}_l(r,r') \\ 
  = ikj_l (kr)(j_l (kr') - iy_l (kr')),r \le r' \\ 
 \end{array}
\end{equation}
\begin{equation}
\begin{array}{l}
 g{}_l(r,r') \\ 
  = ik(j_l (kr) - iy_l (kr))j_l (kr'),r \ge r' \\ 
 \end{array}
\end{equation}
Substitute the $G_L (r,r') $ in (70-72) into the integral equation (69), the 
equation (69) becomes to the following integral equation, 
\begin{equation}
\begin{array}{l}
 \frac{1}{{\varepsilon _\theta  }}E(\vec r') = E^b (\vec r') \\ 
  - r'\sum\limits_{l = 1}^\infty  {\int\limits_{S(r \le R_2 )} {\left( {\frac{\partial }{{\partial r}}\frac{1}{{\varepsilon _\theta  }}} \right)\frac{\partial }{{\partial r}}rg{}_l(r,r')} }  \\ 
 \sum\limits_{m =  - l}^l {} Y_l^{m*} (\theta ,\phi )EdVY_l^m (\theta ',\phi ') \\ 
  + r'\sum\limits_{l = 1}^\infty  {\int\limits_{S(r \le R_2 )} {\frac{1}{{r^2 }}\left( {\frac{1}{{\varepsilon _\theta  }} - \frac{1}{{\varepsilon _r }}} \right)l(l + 1)rg{}_l(r,r')} }  \\ 
 \sum\limits_{m =  - l}^l {Y_l^{m*} (\theta ,\phi )EdVY_l^m (\theta ',\phi ')}  \\ 
  + r'\sum\limits_{l = 1}^\infty  {\int\limits_{S(r \le R_2 )} {k^2 \left( {\frac{1}{{\varepsilon _\theta  }} - \mu _\theta  } \right)rg{}_l(r,r')} }  \\ 
 \sum\limits_{m =  - l}^l {Y_l^{m*} (\theta ,\phi )EdVY_l^m (\theta ',\phi '),}  \\ 
 \end{array}
\end{equation}

\begin{equation}
\begin{array}{l}
 \mathop {\lim }\limits_{r' \to 0} \frac{1}{{\varepsilon _\theta  }}E(\vec r') = \mathop {\lim }\limits_{r' \to 0} E^b (\vec r') \\ 
  - \mathop {\lim }\limits_{r' \to 0} r'\sum\limits_{l = 1}^\infty  {\int\limits_{S(r \le R_2 )} {\left( {\frac{\partial }{{\partial r}}\frac{1}{{\varepsilon _\theta  }}} \right)\frac{\partial }{{\partial r}}rg{}_l(r,r')} }  \\ 
 \sum\limits_{m =  - l}^l {Y_l^{m*} (\theta ,\phi )EdVY_l^m (\theta ',\phi ')}  \\ 
  + \mathop {\lim }\limits_{r' \to 0} r'\sum\limits_{l = 1}^\infty  {\int\limits_{S(r \le R_2 )} {\frac{1}{{r^2 }}\left( {\frac{1}{{\varepsilon _\theta  }} - \frac{1}{{\varepsilon _r }}} \right)l(l + 1)} }  \\ 
 rg{}_l(r,r')\sum\limits_{m =  - l}^l {Y_l^{m*} (\theta ,\phi )EdVY_l^m (\theta ',\phi ')}  \\ 
  + \mathop {\lim }\limits_{r' \to 0} r'\sum\limits_{l = 1}^\infty  {\int\limits_{S(r \le R_2 )} {k^2 \frac{1}{{\varepsilon _\theta  }}rg{}_l(r,r')} }  \\ 
 \sum\limits_{m =  - l}^l {Y_l^{m*} (\theta ,\phi )EdVY_l^m (\theta ',\phi ')}  \\ 
  - \mathop {\lim }\limits_{r' \to 0} r'\sum\limits_{l = 1}^\infty  {\int\limits_{S(r \le R_2 )} {k^2 \mu _\theta  rg{}_l(r,r')} }  \\ 
 \sum\limits_{m =  - l}^l {Y_l^{m*} (\theta ,\phi )EdVY_l^m (\theta ',\phi ')}  \\ 
  = \mathop {\lim }\limits_{r' \to 0} I + \mathop {\lim }\limits_{r' \to 0} II + \mathop {\lim }\limits_{r' \to 0} III \\ 
  + \mathop {\lim }\limits_{r' \to 0} IV + \mathop {\lim }\limits_{r' \to 0} V, \\ 
 \end{array}
\end{equation}

From the equation (60)
\begin{equation}
\mathop {\lim }\limits_{r' \to 0} \frac{1}{{\varepsilon _\theta  }}E(\vec r') = 0
\end{equation}
From the equation (21)
\begin{equation}
\mathop {\lim }\limits_{r' \to 0} I = \mathop {\lim }\limits_{r' \to 0} E^b (\vec r') = 0,
\end{equation}

In next, we will prove 

\begin{equation}
\begin{array}{l}
 \mathop {\lim }\limits_{r' \to 0} II =  \\ 
  =  - \mathop {\lim }\limits_{r' \to 0} r'\sum\limits_{l = 1}^\infty  {\int\limits_{S(r \le R_2 )} {\left( {\frac{\partial }{{\partial r}}\frac{1}{{\varepsilon _\theta  }}} \right)\frac{\partial }{{\partial r}}rg{}_l(r,r')} }  \\ 
 \sum\limits_{m =  - l}^l {Y_l^{m*} (\theta ,\phi )EdVY_l^m (\theta ',\phi ') = 0} , \\ 
 \end{array}
\end{equation}

\begin{equation}
\begin{array}{l}
 \mathop {\lim }\limits_{r' \to 0} III =  \\ 
  = \mathop {\lim }\limits_{r' \to 0} r'\sum\limits_{l = 1}^\infty  {\int\limits_{S(r \le R_2 )} {\left( {\frac{1}{{\varepsilon _\theta  }} - \frac{1}{{\varepsilon _r }}} \right)} } \frac{{l(l + 1)}}{r}g{}_l(r,r') \\ 
 \sum\limits_{m =  - l}^l {} Y_l^{m*} (\theta ,\phi )EdVY_l^m (\theta ',\phi ') = 0, \\ 
 \end{array}
\end{equation}

\begin{equation}
\begin{array}{l}
 \mathop {\lim }\limits_{r' \to 0} IV =  \\ 
  = \mathop {\lim }\limits_{r' \to 0} r'\sum\limits_{l = 1}^\infty  {} \int\limits_{S(r \le R_2 )} {} k^2 \frac{1}{{\varepsilon _\theta  }}rg{}_l(r,r') \\ 
 \sum\limits_{m =  - l}^l {} Y_l^{m*} (\theta ,\phi )EdVY_l^m (\theta ',\phi ') = 0, \\ 
 \end{array}
\end{equation}
\hfill\break
Substitute (70) and (71) into the (77) and using LHOPITAL ROLE, we prove the limitation equation (77) in detail,

\begin{equation}
\begin{array}{l}
 \mathop {\lim }\limits_{r' \to 0} II =  \\ 
  - \mathop {\lim }\limits_{r' \to 0} k\sum\limits_{l = 1}^\infty  {} r'y_l (kr')\int_0^{r'} {} \left( {\frac{\partial }{{\partial r}}\frac{1}{{\varepsilon _\theta  }}} \right)\frac{\partial }{{\partial r}}rj_l (kr)dr \\ 
 \sum\limits_{m =  - l}^l {} \int_0^\pi  {} \int_0^{2\pi } {} Y_l^{m*} (\theta ,\phi )E(r,\theta ,\phi )\sin \theta d\theta d\phi Y_l^m (\theta ',\phi ') \\ 
  - \mathop {\lim }\limits_{r' \to 0} k\sum\limits_{l = 1}^\infty  {} r'j_l (kr')\int_{r'}^{R_2 } {} \left( {\frac{\partial }{{\partial r}}\frac{1}{{\varepsilon _\theta  }}} \right)\frac{\partial }{{\partial r}}ry_l (kr)dr \\ 
 \sum\limits_{m =  - l}^l {} \int_0^\pi  {} \int_0^{2\pi } {} Y_l^{m*} (\theta ,\phi )E(r,\theta ,\phi )\sin \theta d\theta d\phi Y_l^m (\theta ',\phi ') \\ 
  - \mathop {\lim }\limits_{r' \to 0} ik\sum\limits_{l = 1}^\infty  {} r'j_l (kr')\int_0^{R_2 } {} \left( {\frac{\partial }{{\partial r}}\frac{1}{{\varepsilon _\theta  }}} \right)\frac{\partial }{{\partial r}}rj_l (kr)dr \\ 
 \sum\limits_{m =  - l}^l {} \int_0^\pi  {} \int_0^{2\pi } {} Y_l^{m*} (\theta ,\phi )E(r,\theta ,\phi )\sin \theta d\theta d\phi Y_l^m (\theta ',\phi ')\\
 \end{array}
\end{equation}
       
\begin{equation}
\begin{array}{l}
 \mathop {\lim }\limits_{r' \to 0} II =  - \mathop {\lim }\limits_{r' \to 0} k\sum\limits_{l = 1}^\infty  {} r'y_l (kr')\int_0^{r'} {} \left( {\frac{\partial }{{\partial r}}\frac{1}{{\varepsilon _\theta  }}} \right)(l + 1)j_l (kr)dr \\ 
 \sum\limits_{m =  - l}^l {} \int_0^\pi  {} \int_0^{2\pi } {} Y_l^{m*} (\theta ,\phi )E(r,\theta ,\phi )\sin \theta d\theta d\phi Y_l^m (\theta ',\phi ') \\ 
  + \mathop {\lim }\limits_{r' \to 0} k\sum\limits_{l = 1}^\infty  {} r'y_l (kr')\int_0^{r'} {} \left( {\frac{\partial }{{\partial r}}\frac{1}{{\varepsilon _\theta  }}} \right)krj_{l + 1} (kr)dr \\ 
 \sum\limits_{m =  - l}^l {} \int_0^\pi  {} \int_0^{2\pi } {} Y_l^{m*} (\theta ,\phi )E(r,\theta ,\phi )\sin \theta d\theta d\phi Y_l^m (\theta ',\phi ') \\ 
  - \mathop {\lim }\limits_{r' \to 0} k\sum\limits_{l = 1}^\infty  {} r'j_l (kr')\int_{r'}^{R_2 } {} \left( {\frac{\partial }{{\partial r}}\frac{1}{{\varepsilon _\theta  }}} \right)(l + 1)y_l (kr)dr \\ 
 \sum\limits_{m =  - l}^l {} \int_0^\pi  {} \int_0^{2\pi } {} Y_l^{m*} (\theta ,\phi )E(r,\theta ,\phi )\sin \theta d\theta d\phi Y_l^m (\theta ',\phi ') \\ 
  + \mathop {\lim }\limits_{r' \to 0} k\sum\limits_{l = 1}^\infty  {} r'j_l (kr')\int_{r'}^{R_2 } {} \left( {\frac{\partial }{{\partial r}}\frac{1}{{\varepsilon _\theta  }}} \right)kry_{l + 1} (kr)dr \\ 
 \sum\limits_{m =  - l}^l {} \int_0^\pi  {} \int_0^{2\pi } {} Y_l^{m*} (\theta ,\phi )E(r,\theta ,\phi )\sin \theta d\theta d\phi Y_l^m (\theta ',\phi ') \\ 
  - \mathop {\lim }\limits_{r' \to 0} ik\sum\limits_{l = 1}^\infty  {} r'j_l (kr')\int_{r'}^{R_2 } {} \left( {\frac{\partial }{{\partial r}}\frac{1}{{\varepsilon _\theta  }}} \right)\frac{\partial }{{\partial r}}rj_l (kr)dr \\ 
 \sum\limits_{m =  - l}^l {} \int_0^\pi  {} \int_0^{2\pi } {} Y_l^{m*} (\theta ,\phi )E(r,\theta ,\phi )\sin \theta d\theta d\phi Y_l^m (\theta ',\phi ') \\ 
 \end{array}
\end{equation}

By LHOPITAL ROLE

\begin{equation}
\begin{array}{l}
 \mathop {\lim }\limits_{r' \to 0} II =  \\ 
 \mathop {\lim }\limits_{r' \to 0} \sum\limits_{l = 1}^\infty  {} \frac{{(2l)!}}{{2^l l!}}\frac{1}{{kl(kr')^{l - 1} }}\left( {\frac{\partial }{{\partial r}}\frac{1}{{\varepsilon _\theta  }}} \right)(l + 1)\frac{{2^l l!(kr')^l }}{{(2l + 1)!}} \\ 
 \sum\limits_{m =  - l}^l {} \int_0^\pi  {} \int_0^{2\pi } {} Y_l^{m*} (\theta ,\phi )E(r,\theta ,\phi )\sin \theta d\theta d\phi Y_l^m (\theta ',\phi ') \\ 
  - \mathop {\lim }\limits_{r' \to 0} \sum\limits_{l = 1}^\infty  {} \frac{{(2l)!}}{{2^l l!}}\frac{1}{{kl(kr')^{l - 1} }}\left( {\frac{\partial }{{\partial r}}\frac{1}{{\varepsilon _\theta  }}} \right)\frac{{2^{l + 1} (l + 1)!}}{{(2l + 3)!}}(kr')^{l + 2}  \\ 
 \sum\limits_{m =  - l}^l {} \int_0^\pi  {} \int_0^{2\pi } {} Y_l^{m*} (\theta ,\phi )E(r,\theta ,\phi )\sin \theta d\theta d\phi Y_l^m (\theta ',\phi ') \\ 
  - \mathop {\lim }\limits_{r' \to 0} \sum\limits_{l = 1}^\infty  {} \frac{{2^l l!}}{{(2l + 1)!}}\frac{{(kr')^{l + 2} }}{{k(l + 1)}}\left( {\frac{\partial }{{\partial r}}\frac{1}{{\varepsilon _\theta  }}} \right)l\frac{{(2l)!}}{{2^l l!}}\frac{1}{{(kr')^{l + 1} }} \\ 
 \sum\limits_{m =  - l}^l {} \int_0^\pi  {} \int_0^{2\pi } {} Y_l^{m*} (\theta ,\phi )E(r,\theta ,\phi )\sin \theta d\theta d\phi Y_l^m (\theta ',\phi ') \\ 
  - \mathop {\lim }\limits_{r' \to 0} ik\sum\limits_{l = 1}^\infty  {} r'j_l (kr')\int_0^{R_2 } {} \left( {\frac{\partial }{{\partial r}}\frac{1}{{\varepsilon _\theta  }}} \right)\frac{\partial }{{\partial r}}rj_l (kr)dr \\ 
 \sum\limits_{m =  - l}^l {} \int_0^\pi  {} \int_0^{2\pi } {} Y_l^{m*} (\theta ,\phi )E(r,\theta ,\phi )\sin \theta d\theta d\phi Y_l^m (\theta ',\phi '). \\ 
 \end{array}
\end{equation}

\begin{equation}
\begin{array}{l}
 \mathop {\lim }\limits_{r' \to 0} II =  \\ 
 \mathop {\lim }\limits_{r' \to 0} \sum\limits_{l = 1}^\infty  {} \frac{{l + 1}}{{(2l + 1)}}\frac{{r'}}{l}\left( {\frac{\partial }{{\partial r}}\frac{1}{{\varepsilon _\theta  }}} \right) \\ 
 \sum\limits_{m =  - l}^l {} \int_0^\pi  {} \int_0^{2\pi } {} Y_l^{m*} (\theta ,\phi )E(r,\theta ,\phi )\sin \theta d\theta d\phi Y_l^m (\theta ',\phi ') \\ 
  - \mathop {\lim }\limits_{r' \to 0} \sum\limits_{l = 1}^\infty  {} \frac{{(kr')^2 }}{{kl(2l + 3)(2l + 1)}}\left( {\frac{\partial }{{\partial r}}\frac{1}{{\varepsilon _\theta  }}} \right) \\ 
 \sum\limits_{m =  - l}^l {} \int_0^\pi  {} \int_0^{2\pi } {} Y_l^{m*} (\theta ,\phi )E(r,\theta ,\phi )\sin \theta d\theta d\phi Y_l^m (\theta ',\phi ') \\ 
  - \mathop {\lim }\limits_{r' \to 0} \sum\limits_{l = 1}^\infty  {} \frac{l}{{(2l + 1)}}\frac{{r'}}{{(l + 1)}}\left( {\frac{\partial }{{\partial r}}\frac{1}{{\varepsilon _\theta  }}} \right) \\ 
 \sum\limits_{m =  - l}^l {} \int_0^\pi  {} \int_0^{2\pi } {} Y_l^{m*} (\theta ,\phi )E(r,\theta ,\phi )\sin \theta d\theta d\phi Y_l^m (\theta ',\phi ') \\ 
  - \mathop {\lim }\limits_{r' \to 0} ik\sum\limits_{l = 1}^\infty  {} r'j_l (kr')\int_0^{R_2 } {} \left( {\frac{\partial }{{\partial r}}\frac{1}{{\varepsilon _\theta  }}} \right)\frac{\partial }{{\partial r}}rj_l (kr)dr \\ 
 \sum\limits_{m =  - l}^l {} \int_0^\pi  {} \int_0^{2\pi } {} Y_l^{m*} (\theta ,\phi )E(r,\theta ,\phi )\sin \theta d\theta d\phi Y_l^m (\theta ',\phi '), \\ 
 \end{array}    
\end{equation}

             Because the finite energy condition (60) and GLHUA pre cloak material condition $(6.1)\ to \ (6.4)$, the integral in above equation (83) is finite integrative,

\begin{equation}
\begin{array}{l}
 \mathop {\lim }\limits_{r' \to 0} II =  \\ 
 \mathop {\lim }\limits_{r' \to 0} \sum\limits_{l = 1}^\infty  {} \frac{{l + 1}}{{(2l + 1)}}\frac{{r'}}{l}\left( {\frac{\partial }{{\partial r}}\frac{1}{{\varepsilon _\theta  }}} \right) \\ 
 \sum\limits_{m =  - l}^l {} \int_0^\pi  {} \int_0^{2\pi } {} Y_l^{m*} (\theta ,\phi )E(r,\theta ,\phi )\sin \theta d\theta d\phi Y_l^m (\theta ',\phi ') \\ 
  - \mathop {\lim }\limits_{r' \to 0} \sum\limits_{l = 1}^\infty  {} \frac{{(kr')^2 }}{{kl(2l + 3)(2l + 1)}}\left( {\frac{\partial }{{\partial r}}\frac{1}{{\varepsilon _\theta  }}} \right) \\ 
 \sum\limits_{m =  - l}^l {} \int_0^\pi  {} \int_0^{2\pi } {} Y_l^{m*} (\theta ,\phi )E(r,\theta ,\phi )\sin \theta d\theta d\phi Y_l^m (\theta ',\phi ') \\ 
  - \mathop {\lim }\limits_{r' \to 0} \sum\limits_{l = 1}^\infty  {} \frac{l}{{(2l + 1)}}\frac{{r'}}{{(l + 1)}}\left( {\frac{\partial }{{\partial r}}\frac{1}{{\varepsilon _\theta  }}} \right) \\ 
 \sum\limits_{m =  - l}^l {} \int_0^\pi  {} \int_0^{2\pi } {} Y_l^{m*} (\theta ,\phi )E(r,\theta ,\phi )\sin \theta d\theta d\phi Y_l^m (\theta ',\phi ') \\ 
  - \mathop {\lim }\limits_{r' \to 0} ik\sum\limits_{l = 1}^\infty  {} r'j_l (kr')\int_0^{R_2 } {} \left( {\frac{\partial }{{\partial r}}\frac{1}{{\varepsilon _\theta  }}} \right)\frac{\partial }{{\partial r}}rj_l (kr)dr \\ 
 \sum\limits_{m =  - l}^l {} \int_0^\pi  {} \int_0^{2\pi } {} Y_l^{m*} (\theta ,\phi )E(r,\theta ,\phi )\sin \theta d\theta d\phi Y_l^m (\theta ',\phi ') \\ 
  = 0, \\ 
 \end{array}
\end{equation}

Therefore,

\begin{equation}
\begin{array}{l}
 \mathop {\lim }\limits_{r' \to 0} II =  - \mathop {\lim }\limits_{r' \to 0} r'\sum\limits_{l = 1}^\infty  {} \int\limits_{S(r \le R_2 )} {} \left( {\frac{\partial }{{\partial r}}\frac{1}{{\varepsilon _\theta  }}} \right)\frac{\partial }{{\partial r}}rg{}_l(r,r') \\ 
 \sum\limits_{m =  - l}^l {} Y_l^{m*} (\theta ,\phi )EdVY_l^m (\theta ',\phi ') = 0, \\ 
 \end{array}
\end{equation}

limitation equation (77) is proved. Similarly, we can prove limitation equation (78)

\begin{equation}
\begin{array}{l}
 \mathop {\lim }\limits_{r' \to 0} III = \mathop {\lim }\limits_{r' \to 0} r'\sum\limits_{l = 1}^\infty  {} \int\limits_{S(r \le R_2 )} {} \frac{1}{{r^2 }}\left( {\frac{1}{{\varepsilon _\theta  }} - \frac{1}{{\varepsilon _r }}} \right)l(l + 1)rg{}_l(r,r') \\ 
 \sum\limits_{m =  - l}^l {} Y_l^{m*} (\theta ,\phi )EdVY_l^m (\theta ',\phi ') = 0, \\ 
 \end{array}
\end{equation}

,
and (79),

\begin{equation}
\begin{array}{l}
 \mathop {\lim }\limits_{r' \to 0} IV = \mathop {\lim }\limits_{r' \to 0} r'\sum\limits_{l = 1}^\infty  {} \int\limits_{S(r \le R_2 )} {} k^2 \frac{1}{{\varepsilon _\theta  }}rg{}_l(r,r') \\ 
 \sum\limits_{m =  - l}^l {} Y_l^{m*} (\theta ,\phi )EdVY_l^m (\theta ',\phi ') = 0, \\ 
 \end{array}
\end{equation}
Substitute (75)-(79) into the limitation equation (74), the limitation equation (74) induces

\begin{equation}
\begin{array}{l}
 \mathop {\lim }\limits_{r' \to 0} V =  - \mathop {\lim }\limits_{r' \to 0} r'\sum\limits_{l = 1}^\infty  {\int\limits_{S(r \le R_2 )} {} k^2 \mu _\theta  rg{}_l(r,r')}  \\ 
 \sum\limits_{m =  - l}^l {Y_l^{m*} (\theta ,\phi )EdVY_l^m (\theta ',\phi ')}  = 0, \\ 
 \end{array}
\end{equation}
Substitute (71) and (72) for $g{}_l(r,r')$ into the (88),
\begin{equation}
\begin{array}{l}
 \mathop {\lim }\limits_{r' \to 0} V =  \\ 
  - \mathop {\lim }\limits_{r' \to 0} k^3 \sum\limits_{l = 1}^\infty  {} r'y_l (kr')\int_0^{r'} {} \mu _\theta  rj_l (kr)dr \\ 
 \sum\limits_{m =  - l}^l {} \int_0^\pi  {} \int_0^{2\pi } {} Y_l^{m*} (\theta ,\phi )E(r,\theta ,\phi )\sin \theta d\theta d\phi Y_l^m (\theta ',\phi ') \\ 
  - \mathop {\lim }\limits_{r' \to 0} k^3 \sum\limits_{l = 1}^\infty  {} r'j_l (kr')\int_{r'}^{R_2 } {} \mu _\theta  ry_l (kr)dr \\ 
 \sum\limits_{m =  - l}^l {} \int_0^\pi  {} \int_0^{2\pi } {} Y_l^{m*} (\theta ,\phi )E(r,\theta ,\phi )\sin \theta d\theta d\phi Y_l^m (\theta ',\phi ') \\ 
  - \mathop {\lim }\limits_{r' \to 0} ik^3 \sum\limits_{l = 1}^\infty  {} r'j_l (kr')\int_0^{R_2 } {} \mu _\theta  rj_l (kr)dr \\ 
 \sum\limits_{m =  - l}^l {} \int_0^\pi  {} \int_0^{2\pi } {} Y_l^{m*} (\theta ,\phi )E(r,\theta ,\phi )\sin \theta d\theta d\phi Y_l^m (\theta ',\phi ') \\ 
  = 0 \\ 
 \end{array}
\end{equation}

\begin{equation}
\begin{array}{l}
 \mathop {\lim }\limits_{r' \to 0} V =  \\ 
  + \mathop {\lim }\limits_{r' \to 0} k^2 \sum\limits_{l = 1}^\infty  {} \frac{{(2l)!}}{{2^l l!}}\frac{1}{{(kr')^l }}\int_0^{r'} {} \mu _\theta  rj_l (kr)dr \\ 
 \sum\limits_{m =  - l}^l {} \int_0^\pi  {} \int_0^{2\pi } {} Y_l^{m*} (\theta ,\phi )E(r,\theta ,\phi )\sin \theta d\theta d\phi Y_l^m (\theta ',\phi ') \\ 
  - \mathop {\lim }\limits_{r' \to 0} k^2 \sum\limits_{l = 1}^\infty  {} \frac{{2^l l!(kr')^{l + 1} }}{{(2l + 1)!}}\int_{r'}^{R_2 } {} \mu _\theta  ry_l (kr)dr \\ 
 \sum\limits_{m =  - l}^l {} \int_0^\pi  {} \int_0^{2\pi } {} Y_l^{m*} (\theta ,\phi )E(r,\theta ,\phi )\sin \theta d\theta d\phi Y_l^m (\theta ',\phi ') \\ 
  - \mathop {\lim }\limits_{r' \to 0} ik^3 \sum\limits_{l = 1}^\infty  {} r'j_l (kr')\int_0^{R_2 } {} \mu _\theta  rj_l (kr)dr \\ 
 \sum\limits_{m =  - l}^l {} \int_0^\pi  {} \int_0^{2\pi } {} Y_l^{m*} (\theta ,\phi )E(r,\theta ,\phi )\sin \theta d\theta d\phi Y_l^m (\theta ',\phi ') \\ 
  = 0, \\ 
 \end{array}
\end{equation}

By LHOPITAL ROLE

\begin{equation}
\begin{array}{l}
 \mathop {\lim }\limits_{r' \to 0} V = \mathop {\lim }\limits_{r' \to 0} k^2 \sum\limits_{l = 1}^\infty  {} \frac{{(2l)!}}{{2^l l!}}\frac{1}{{kl(kr')^{l - 1} }}\mu _\theta  r'j_l (kr') \\ 
 \sum\limits_{m =  - l}^l {} \int_0^\pi  {} \int_0^{2\pi } {} Y_l^{m*} (\theta ,\phi )E(r,\theta ,\phi )\sin \theta d\theta d\phi Y_l^m (\theta ',\phi ') \\ 
  - \mathop {\lim }\limits_{r' \to 0} k^2 \sum\limits_{l = 1}^\infty  {} \frac{{2^l l!(kr')^{l + 2} }}{{k(l + 1)(2l + 1)!}}\mu _\theta  r'y_l (kr') \\ 
 \sum\limits_{m =  - l}^l {} \int_0^\pi  {} \int_0^{2\pi } {} Y_l^{m*} (\theta ,\phi )E(r,\theta ,\phi )\sin \theta d\theta d\phi Y_l^m (\theta ',\phi ') \\ 
  - \mathop {\lim }\limits_{r' \to 0} ik^3 \sum\limits_{l = 1}^\infty  {} r'j_l (kr')\int_0^{R_2 } {} \mu _\theta  rj_l (kr)dr \\ 
 \sum\limits_{m =  - l}^l {} \int_0^\pi  {} \int_0^{2\pi } {} Y_l^{m*} (\theta ,\phi )E(r,\theta ,\phi )\sin \theta d\theta d\phi Y_l^m (\theta ',\phi ') \\ 
  = 0, \\ 
 \end{array}
\end{equation}

\begin{equation}
\begin{array}{l}
 \mathop {\lim }\limits_{r' \to 0} V =  \\ 
 \mathop {\lim }\limits_{r' \to 0} k^2 \sum\limits_{l = 1}^\infty  {} \frac{{(2l)!}}{{2^l l!}}\frac{1}{{kl(kr')^{l - 1} }}\frac{{f(r')}}{{r'}}\frac{{2^l l!}}{{(2l + 1)!}}(kr')^l  \\ 
 \sum\limits_{m =  - l}^l {} \int_0^\pi  {} \int_0^{2\pi } {} Y_l^{m*} (\theta ,\phi )E(r,\theta ,\phi )\sin \theta d\theta d\phi Y_l^m (\theta ',\phi ') \\ 
  + \mathop {\lim }\limits_{r' \to 0} k^2 \sum\limits_{l = 1}^\infty  {} \frac{{2^l l!(kr')^{l + 2} }}{{k(l + 1)(2l + 1)!}}\frac{{f(r')}}{{r'}}\frac{{(2l)!}}{{2^l l!}}\frac{1}{{(kr')^{l + 1} }} \\ 
 \sum\limits_{m =  - l}^l {} \int_0^\pi  {} \int_0^{2\pi } {} Y_l^{m*} (\theta ,\phi )E(r,\theta ,\phi )\sin \theta d\theta d\phi Y_l^m (\theta ',\phi ') \\ 
  - \mathop {\lim }\limits_{r' \to 0} ik^3 \sum\limits_{l = 1}^\infty  {} r'j_l (kr')\int_0^{R_2 } {} \varepsilon _\theta  rj_l (kr)dr \\ 
 \sum\limits_{m =  - l}^l {} \int_0^\pi  {} \int_0^{2\pi } {} Y_l^{m*} (\theta ,\phi )E(r,\theta ,\phi )\sin \theta d\theta d\phi Y_l^m (\theta ',\phi ') \\ 
  = 0, \\ 
 \end{array}
\end{equation}

\begin{equation}
\begin{array}{l}
 \mathop {\lim }\limits_{r' \to 0} V =  \\ 
 \mathop {\lim }\limits_{r' \to 0} k^2 f(r')\sum\limits_{l = 1}^\infty  {} \frac{1}{{l(2l + 1)}} \\ 
 \sum\limits_{m =  - l}^l {} \int_0^\pi  {} \int_0^{2\pi } {} Y_l^{m*} (\theta ,\phi )E(r,\theta ,\phi )\sin \theta d\theta d\phi Y_l^m (\theta ',\phi ') \\ 
  + \mathop {\lim }\limits_{r' \to 0} k^2 f(r')\sum\limits_{l = 1}^\infty  {} \frac{1}{{(l + 1)(2l + 1)}} \\ 
 \sum\limits_{m =  - l}^l {} \int_0^\pi  {} \int_0^{2\pi } {} Y_l^{m*} (\theta ,\phi )E(r,\theta ,\phi )\sin \theta d\theta d\phi Y_l^m (\theta ',\phi ') \\ 
  - \mathop {\lim }\limits_{r' \to 0} ik^3 \sum\limits_{l = 1}^\infty  {} r'j_l (kr')\int_0^{R_2 } {} \varepsilon _\theta  rj_l (kr)dr \\ 
 \sum\limits_{m =  - l}^l {} \int_0^\pi  {} \int_0^{2\pi } {} Y_l^{m*} (\theta ,\phi )E(r,\theta ,\phi )\sin \theta d\theta d\phi Y_l^m (\theta ',\phi ') \\ 
  = 0, \\ 
 \end{array}
\end{equation}

\begin{equation}
\begin{array}{l}
 \mathop {\lim }\limits_{r' \to 0} k^2 R_2 ^2 \sum\limits_{l = 1}^\infty  {} \frac{1}{{l(l + 1)}} \\ 
 \sum\limits_{m =  - l}^l {} \int_0^\pi  {} \int_0^{2\pi } {} Y_l^{m*} (\theta ,\phi )E(r,\theta ,\phi )\sin \theta d\theta d\phi  \\ 
 Y_l^m (\theta ',\phi ') = 0, \\ 
 \end{array}
\end{equation}
Because for incident wave with electric current point source $\delta (\vec r - \vec r_s )\vec e_s $ at $(r_s ,\theta _s ,\phi _s )$
$r_s  > R_O  > R_2 $, $E_0 (r) = 0$ in (30),

\begin{equation}
\begin{array}{l}
 E(r',\theta ',\phi ') = \sum\limits_{l = 1}^\infty  {} E_l (r') \\ 
 \sum\limits_{m =  - l}^l {} Y_l^m (\theta ',\phi ')Y_l^m (\theta _s ,\phi _s ) \\ 
 \end{array}
\end{equation}

Substitute (95) into the (94), we have

\begin{equation}
\begin{array}{l}
 \mathop {\lim }\limits_{r' \to 0} k^2 R_2 ^2 \sum\limits_{l = 1}^\infty  {} \frac{1}{{l(l + 1)}}E_l (r') \\ 
 \sum\limits_{m =  - l}^l {} Y_l^m (\theta ',\phi ')Y_l^m (\theta _s ,\phi _s ) = 0, \\ 
 \end{array}
\end{equation}

Let

\begin{equation}
\Re (\theta ,\phi ) = \frac{1}{{\sin \theta }}\frac{\partial }{{\partial \theta }}\sin \theta \frac{\partial }{{\partial \theta }} + \frac{1}{{\sin ^2 \theta }}\frac{{\partial ^2 }}{{\partial \phi ^2 }},
\end{equation}

To use $\Re (\theta ,\phi )$ in (97) to make action to both side of the limitation equation (96), we have

\begin{equation}
\begin{array}{l}
 \mathop {\lim }\limits_{r' \to 0} k^2 R_2 ^2 \Re (\theta ',\phi ')\sum\limits_{l = 1}^\infty  {} \frac{1}{{l(l + 1)}}E_l (r') \\ 
 \sum\limits_{m =  - l}^l {} Y_l^m (\theta ',\phi ')Y_l^m (\theta _s ,\phi _s ) = 0 \\ 
 \end{array}
\end{equation}

\begin{equation}
\mathop {\lim }\limits_{r' \to 0} k^2 R_2 ^2 E(r',\theta ',\phi ') = 0,
\end{equation}

For every $(\theta ',\phi ')$ and $k \ge k_0  > 0$

\begin{equation}
\mathop {\lim }\limits_{r' \to 0} E(r',\theta ',\phi ') = 0,
\end{equation}

Other proof approach is as follows: By using above similar with proof process of (98), for every term $l$ and $k$, we can prove that

\begin{equation}
\mathop {\lim }\limits_{r' \to 0} k^2 R_2 ^2 \frac{1}{{l(l + 1)}}E_l (r') = 0,
\end{equation}
,                     
Because $E_0 (r) = 0$ in (30),$l \ge 1$ and for $k \ge k_0  > 0$
\begin{equation}
\mathop {\lim }\limits_{r' \to 0} E_l (r') = 0,
\end{equation}

\begin{equation}
\begin{array}{l}
 \mathop {\lim }\limits_{r' \to 0} E(r',\theta ',\phi ') = \sum\limits_{l = 1}^\infty  {} \mathop {\lim }\limits_{r' \to 0} E_l (r') \\ 
 \sum\limits_{m =  - l}^l {} Y_l^m (\theta ',\phi ')Y_l^m (\theta _s ,\phi _s ) = 0, \\ 
 \end{array}
\end{equation}

The limitation equation (61) is proved. Similarly, we can prove limitation equation (62)

\begin{equation}
\begin{array}{l}
 \mathop {\lim }\limits_{r' \to 0} H(r',\theta ',\phi ') = \sum\limits_{l = 1}^\infty  {} \mathop {\lim }\limits_{r' \to 0} H_l (r') \\ 
 \sum\limits_{m =  - l}^l {} Y_l^m (\theta ',\phi ')Y_l^m (\theta _s ,\phi _s ) = 0, \\ 
 \end{array}
\end{equation}

Similarly, for incident plane electromagnetic wave , we also can prove limitation equation (61)
and (62). The theorem 6.1 is proved.
\hfill\break

${\boldsymbol{Theorem \ 6.2 }}$,\ Suppose that the anisotropic relative electric permittivity $\varepsilon _r \left( r \right)$,$\varepsilon _\theta  \left( r \right)$ $\varepsilon _\phi  \left( r \right)$ and magnetic permeability, $\mu _r \left( r \right)$,$\mu _\theta  \left( r \right)$ $\mu _\phi  \left( r \right)$ , satisfy the above GLHUA pre cloak material conditions in invisible sphere $(6.1) \ to \ (6.4)$, and finite energy condition (60), then

\begin{equation}
\mathop {\lim }\limits_{r \to 0} \frac{1}{{\varepsilon _\theta  }}\frac{\partial }{{\partial r}}E(\vec r) = 0,
\end{equation}
               
\begin{equation} 
\mathop {\lim }\limits_{r \to 0} \frac{1}{{\varepsilon _\theta  }}\frac{\partial }{{\partial r}}H(\vec r) = 0,
\end{equation}
\hfill\break

${\boldsymbol{Proof:}}$,\ Using similar proof process, the theorem 6.2 can be proved.
\hfill\break

${\boldsymbol{Theorem \ 6.3 }}$,\ Suppose that the anisotropic relative electric permittivity $\varepsilon _r \left( r \right)$,$\varepsilon _\theta  \left( r \right)$ $\varepsilon _\phi  \left( r \right)$ and magnetic permeability, $\mu _r \left( r \right)$,$\mu _\theta  \left( r \right)$ $\mu _\phi  \left( r \right)$ , satisfy the above GLHUA pre cloak material conditions in invisible sphere $(6.1) \ to \ (6.4)$, also angular electromagnetic wave satisfy following finite energy condition (63)

\begin{equation}
\begin{array}{l}
 \int\limits_{S(r \le R_2 )} {\left( {\left| {rE_\theta  (\vec r)} \right|^2  + \left| {rH_\theta  (\vec r)} \right|^2 } \right.}  \\ 
 \left. { + \left| {rE_\phi  (\vec r)} \right|^2  + \left| {rH_\phi  (\vec r)} \right|^2 } \right)dV, \\ \ \ is \ finite, \\
 \end{array}
\end{equation}

then                           

\begin{equation}
\begin{array}{l}
 \mathop {\lim }\limits_{r \to 0} rE_\theta  (\vec r) = 0, \\ 
 \mathop {\lim }\limits_{r \to 0} rE_\phi  (\vec r) = 0, \\ 
 \end{array}
\end{equation}
\begin{equation}
\begin{array}{l}
 \mathop {\lim }\limits_{r \to 0} rH_\theta  (\vec r) = 0, \\ 
 \mathop {\lim }\limits_{r \to 0} rH_\phi  (\vec r) = 0, \\ 
 \end{array}
\end{equation}
\hfill\break

 ${\boldsymbol{Proof:}}$,\
Because the source is outside sphere,$r_s  > R_O  > R_2 $, In the Sphere $0 < r \le R_2 $, from (4), we have

\begin{equation}
\begin{array}{l}
 \nabla  \cdot \vec D = \frac{1}{{r^2 }}\frac{\partial }{{\partial r}}\left( {r^2 \varepsilon _r E_r } \right) +  \\ 
  + \frac{1}{{\sin \theta r}}\frac{\partial }{{\partial \theta }}\sin \theta \varepsilon _\theta  E_\theta   + \frac{1}{{r\sin \theta }}\frac{\partial }{{\partial \phi }}\varepsilon _\phi  E_\phi   = 0 \\ 
 \end{array}
\end{equation}
The $\vec D$ is displacement electric in spherical coordinate, by definition of GL electromagnetic wave
(18), equation (110) becomes

\begin{equation}
\begin{array}{l}
 \frac{1}{{\sin \theta }}\frac{\partial }{{\partial \theta }}\sin \theta rE_\theta   + \frac{1}{{\sin \theta }}\frac{\partial }{{\partial \phi }}rE_\phi   \\ 
  =  - \frac{1}{{\varepsilon _\theta  }}\frac{{\partial E}}{{\partial r}}, \\ 
 \end{array}
\end{equation}

By Maxwell equation (1) and (18), we have

\begin{equation}
\begin{array}{l}
  - \frac{1}{{\sin \theta }}\frac{\partial }{{\partial \phi }}rE_\theta   + \frac{1}{{\sin \theta }}\frac{\partial }{{\partial \theta }}\sin \theta rE_\phi   \\ 
  =  - i\omega \mu _0 H, \\ 
 \end{array}
\end{equation}

Rewrite (111) and (112) as matrix equation

\begin{equation}
\begin{array}{l}
 \left[ {\begin{array}{*{20}c}
   {\frac{1}{{\sin \theta }}\frac{\partial }{{\partial \theta }}\sin \theta } & {\frac{1}{{\sin \theta }}\frac{\partial }{{\partial \phi }}}  \\
   { - \frac{1}{{\sin \theta }}\frac{\partial }{{\partial \phi }}} & {\frac{1}{{\sin \theta }}\frac{\partial }{{\partial \theta }}\sin \theta }  \\
\end{array}} \right]\left[ {\begin{array}{*{20}c}
   {rE_\theta  }  \\
   {rE_\phi  }  \\
\end{array}} \right] =  \\ 
  = \left[ {\begin{array}{*{20}c}
   { - \frac{1}{{\varepsilon _\theta  }}\frac{{\partial E}}{{\partial r}}}  \\
   { - i\omega \mu _0 H}  \\
\end{array}} \right] \\ 
 \end{array}
\end{equation}

The adjoint Greens equation of equation (113) on $[0,\pi ;0,2\pi ]$ is

\begin{equation}
\begin{array}{l}
 \left[ {\begin{array}{*{20}c}
   {\frac{1}{{\sin \theta }}\frac{\partial }{{\partial \theta }}\sin \theta } & { - \frac{1}{{\sin \theta }}\frac{\partial }{{\partial \phi }}}  \\
   {\frac{1}{{\sin \theta }}\frac{\partial }{{\partial \phi }}} & {\frac{1}{{\sin \theta }}\frac{\partial }{{\partial \theta }}\sin \theta }  \\
\end{array}} \right] \\ 
 \left[ {\begin{array}{*{20}c}
   {G_{11} (\theta ,\theta ',\phi ,\phi ')} & {G_{21} (\theta ,\theta ',\phi ,\phi ')}  \\
   {G_{12} (\theta ,\theta ',\phi ,\phi ')} & {G_{22} (\theta ,\theta ',\phi ,\phi ')}  \\
\end{array}} \right] \\ 
  = \frac{1}{{\sin \theta }}\left[ {\begin{array}{*{20}c}
   {\delta (\theta ,\theta ',\phi ,\phi ')} & {}  \\
   {} & {\delta (\theta ,\theta ',\phi ,\phi ')}  \\
\end{array}} \right], \\ 
 \end{array}
\end{equation}

It is different from GL Green equation in (44), the adjoint Greens equation is novel GLHUA angular Green equation. It is different from GL Green function in (45), the GLHUA angular Green function matrix is in the following 

\begin{equation}
\begin{array}{l}
 G_{11} (\theta ,\phi ,\theta ',\phi ') =  \\ 
 \sum\limits_{l = }^\infty  {} \frac{1}{{l(l + 1)}}\sum\limits_{m =  - l}^l {\frac{\partial }{{\partial \theta }}Y_l ^{m*} (\theta ,\phi )} Y_l ^{m*} (\theta ',\phi '), \\ 
 G_{12} (\theta ,\phi ,\theta ',\phi ') =  \\ 
  - \sum\limits_{l = }^\infty  {} \frac{1}{{l(l + 1)}}\sum\limits_{m =  - l}^l {\frac{\partial }{{\sin \theta \partial \phi }}Y_l ^{m*} (\theta ,\phi )} Y_l ^{m*} (\theta ',\phi '), \\ 
 G_{21} (\theta ,\phi ,\theta ',\phi ') =  \\ 
 \sum\limits_{l = }^\infty  {} \frac{1}{{l(l + 1)}}\sum\limits_{m =  - l}^l {\frac{\partial }{{\sin \theta \partial \phi }}Y_l ^{m*} (\theta ,\phi )} Y_l ^{m*} (\theta ',\phi '), \\ 
 G_{22} (\theta ,\phi ,\theta ',\phi ') =  \\ 
 \sum\limits_{l = }^\infty  {} \frac{1}{{l(l + 1)}}\sum\limits_{m =  - l}^l {\frac{\partial }{{\partial \theta }}Y_l ^{m*} (\theta ,\phi )} Y_l ^{m*} (\theta ',\phi '), \\ 
 \end{array}
\end{equation}

To use product of GLHUA Greens function matrix by $\sin \theta $,
$G(\theta ,\theta ',\phi ,\phi ')\sin \theta $, to multiply the matrix equation (113) and  take sphere surface integral of resulted equation on 
$[0,\pi ;0,2\pi ]$

\begin{equation}
\begin{array}{l}
  \\ 
 \int_0^\pi  {} \int_0^{2\pi } {} \left( {\left[ {\begin{array}{*{20}c}
   {G_{11} } & {G_{12} }  \\
   {G_{21} } & {G_{22} }  \\
\end{array}} \right]} \right. \\ 
 \left. {\left[ {\begin{array}{*{20}c}
   {\frac{1}{{\sin \theta }}\frac{\partial }{{\partial \theta }}\sin \theta } & {\frac{1}{{\sin \theta }}\frac{\partial }{{\partial \phi }}}  \\
   { - \frac{1}{{\sin \theta }}\frac{\partial }{{\partial \phi }}} & {\frac{1}{{\sin \theta }}\frac{\partial }{{\partial \theta }}\sin \theta }  \\
\end{array}} \right]\left[ {\begin{array}{*{20}c}
   {rE_\theta  }  \\
   {rE_\phi  }  \\
\end{array}} \right]} \right)\sin \theta d\theta d\phi  \\ 
  = \int_0^\pi  {} \int_0^{2\pi } {} \left[ {\begin{array}{*{20}c}
   {G_{11} } & {G_{12} }  \\
   {G_{21} } & {G_{22} }  \\
\end{array}} \right]\left[ {\begin{array}{*{20}c}
   { - \frac{1}{{\varepsilon _\theta  }}\frac{{\partial E}}{{\partial r}}}  \\
   { - i\omega \mu _0 H}  \\
\end{array}} \right]\sin \theta d\theta d\phi , \\ 
 \end{array}
\end{equation}
 
To use vector $[rE_\theta  ,rE_\phi  ]\sin \theta $ to multiply the adjoint Greens equation of equation (114) and take sphere surface integral of resulted adjoint equation on $[0,\pi ;0,2\pi ]$,

\begin{equation}
\begin{array}{l}
 \int_0^\pi  {} \int_0^{2\pi } {} \left[ {\begin{array}{*{20}c}
   {G_{11} } & {G_{12} }  \\
   {G_{21} } & {G_{22} }  \\
\end{array}} \right] \\ 
 \left[ {\begin{array}{*{20}c}
   {\frac{1}{{\sin \theta }}\frac{\partial }{{\partial \theta }}\sin \theta } & {\frac{1}{{\sin \theta }}\frac{\partial }{{\partial \phi }}}  \\
   { - \frac{1}{{\sin \theta }}\frac{\partial }{{\partial \phi }}} & {\frac{1}{{\sin \theta }}\frac{\partial }{{\partial \theta }}\sin \theta }  \\
\end{array}} \right]\left[ {\begin{array}{*{20}c}
   {rE_\theta  }  \\
   {rE_\phi  }  \\
\end{array}} \right]\sin \theta d\theta d\phi  \\ 
  = \int_0^\pi  {} \int_0^{2\pi } {} \left[ {\begin{array}{*{20}c}
   \delta  & {}  \\
   {} & \delta   \\
\end{array}} \right]\left[ {\begin{array}{*{20}c}
   {rE_\theta  }  \\
   {rE_\phi  }  \\
\end{array}} \right]\sin \theta d\theta d\phi , \\ 
 \end{array}
\end{equation}
To subtract equation (116) from (117), we have
\begin{equation}
\begin{array}{l}
 \left[ {\begin{array}{*{20}c}
   {rE_\theta  (r,\theta ',\phi ')}  \\
   {rE_\phi  (r,\theta ',\phi ')}  \\
\end{array}} \right] =  \\ 
  = \int_0^\pi  {} \int_0^{2\pi } {} \left[ {\begin{array}{*{20}c}
   {G_{11} } & {G_{12} }  \\
   {G_{21} } & {G_{22} }  \\
\end{array}} \right]\left[ {\begin{array}{*{20}c}
   { - \frac{1}{{\varepsilon _\theta  }}\frac{{\partial E}}{{\partial r}}}  \\
   { - i\omega \mu _0 H}  \\
\end{array}} \right]\sin \theta d\theta d\phi  \\ 
 \end{array}
\end{equation}

Based on theorem 6.1 and theorem 6.2, we have
\begin{equation}
\begin{array}{l}
 \mathop {\lim }\limits_{r \to 0} \left[ {\begin{array}{*{20}c}
   {rE_\theta  (r,\theta ',\phi ')}  \\
   {rE_\phi  (r,\theta ',\phi ')}  \\
\end{array}} \right] \\ 
  = \mathop {\lim }\limits_{r \to 0} \int_0^\pi  {} \int_0^{2\pi } {} \left[ {\begin{array}{*{20}c}
   {G_{11} } & {G_{12} }  \\
   {G_{21} } & {G_{22} }  \\
\end{array}} \right]\left[ {\begin{array}{*{20}c}
   { - \frac{1}{{\varepsilon _\theta  }}\frac{{\partial E}}{{\partial r}}}  \\
   { - i\omega \mu _0 H}  \\
\end{array}} \right]\sin \theta d\theta d\phi  \\ 
  = \int_0^\pi  {} \int_0^{2\pi } {} \left[ {\begin{array}{*{20}c}
   {G_{11} } & {G_{12} }  \\
   {G_{21} } & {G_{22} }  \\
\end{array}} \right]\mathop {\lim }\limits_{r \to 0} \left[ {\begin{array}{*{20}c}
   { - \frac{1}{{\varepsilon _\theta  }}\frac{{\partial E}}{{\partial r}}}  \\
   { - i\omega \mu _0 H}  \\
\end{array}} \right]\sin \theta d\theta d\phi  \\ 
  = \left[ {\begin{array}{*{20}c}
   0  \\
   0  \\
\end{array}} \right], \\ 
 \end{array}
\end{equation}

We have proved the first part of (108) of theorem 6.3, similarly, we can prove second part of (109). The theorem 6.3 is proved.
\hfill\break

${\boldsymbol{Theorem \ 6.4 }}$,\ Suppose that the anisotropic relative electric permittivity $\varepsilon _r \left( r \right)$,$\varepsilon _\theta  \left( r \right)$ $\varepsilon _\phi  \left( r \right)$ and magnetic permeability, $\mu _r \left( r \right)$,$\mu _\theta  \left( r \right)$ $\mu _\phi  \left( r \right)$ , satisfy the above GLHUA pre cloak material conditions in invisible sphere $(6.1) \ to \ (6.4)$, also angular electromagnetic wave satisfy following finite energy condition (63), then
\begin{equation}
\begin{array}{l}
 \mathop {\lim }\limits_{r \to 0} \frac{1}{{\varepsilon _\theta  }}\frac{\partial }{{\partial r}}rE_\theta  (\vec r) = 0, \\ 
 \mathop {\lim }\limits_{r \to 0} \frac{1}{{\varepsilon _\theta  }}\frac{\partial }{{\partial r}}rE_\phi  (\vec r) = 0, \\ 
 \end{array}
\end{equation}

\begin{equation}
\begin{array}{l}
 \mathop {\lim }\limits_{r \to 0} \frac{1}{{\mu _\theta  }}\frac{\partial }{{\partial r}}rH_\theta  (\vec r) = 0, \\ 
 \mathop {\lim }\limits_{r \to 0} \frac{1}{{\mu _\theta  }}\frac{\partial }{{\partial r}}rH_\phi  (\vec r) = 0, \\ 
 \end{array}
\end{equation}
\hfill\break

${\boldsymbol{Proof : }}$,\ By using the similar proof process on the 8.3, we can prove the theorem 6.4.
\section{discussion and conclusion}
The pre cloak material conditions $(6.1) \ to \ (6.4)$ in GLHUA sphere is from
GL zero scattering inversion and GL no scattering modeling.
Many GL no scattering modeling simulations show that under the conditions $(6.1) 
\ to \ (6.4)$, $\mathop {\lim }\limits_{r \to 0} E(r) = 0,$, and $\mathop {\lim }\limits_{r \to 0} H(r) = 0,$ are verified.
These condition is not unique and can be relaxed. 
We publish this paper to arXiv for support our paper arXiv.org/abs/1612.02857.
Our GLHUA cloak and GLHUA sphere publication in arXiv are for open review.
Please colleague give comments to me by my email or give open comments in arXiv. 
All copyright and patent of the GLHUA EM cloaks,GLHUA sphere and GL modeling
and inversion
methods are reserved by authors in GL Geophysical Laboratory.If some colleague
cite our paper in his work paper, please cite our paper as reference in his paper..

\begin{acknowledgments}
We wish to acknowledge the support of the GL Geophysical Laboratory and thank the GLGEO Laboratory to approve the paper
publication.
 
\end{acknowledgments}



\begin{thebibliography}{99}
\bibitem{1}
Xie,~G., J.H.~Li, E. ~Majer, D. ~Zuo, M. ~Oristaglio
``3-D electromagnetic modeling and nonlinear inversion,"
\textit{Geophysics}, Vol.~65, No.~3, 804--822, 2000.
\bibitem{2} 
Xie,~G., F.~Xie, L.~Xie, and J.~Li,
``New GL method and its advantages for resolving historical diffculties,"
\textit{Progress In Electromagnetics Research}, PIER~63, 141--152, 2006.
\bibitem{3}
Xie,~G., J.~Li, L.~Xie, and F.~Xie,
``GL metro carlo EM inversion,"
\textit{Journal of Electromagnetic Waves and Applications}, Vol.~20, No.~14, 1991--2000, 2006.
\end{thebibliography}
\end{document}